# Compressed Smooth Sparse Decomposition[*]

Shancong Mou and Jianjun Shi[†]

*H. Milton Stewart School of Industrial and Systems Engineering, Georgia Institute of Technology, Atlanta, GA, 30332, USA.*
shancong.mou@gatech.edu, jianjun.shi@isye.gatech.edu

Image-based anomaly detection systems are of vital importance in various manufacturing applications. The resolution and acquisition rate of such systems is increasing significantly in recent years under the fast development of image sensing technology. This enables the detection of tiny defects in real-time. However, such a high resolution and acquisition rate of image data not only slows down the speed of image processing algorithms but also increases data storage and transmission cost. To tackle this problem, we propose a fast and data-efficient method with theoretical performance guarantee that is suitable for sparse anomaly detection in images with a smooth background (smooth plus sparse signal). The proposed method, named Compressed Smooth Sparse Decomposition (CSSD), is a one-step method that unifies the compressive image acquisition and decomposition-based image processing techniques. To further enhance its performance in a high-dimensional scenario, a Kronecker Compressed Smooth Sparse Decomposition (KronCSSD) method is proposed. Compared to traditional smooth and sparse decomposition algorithms, significant transmission cost reduction and computational speed boost can be achieved with negligible performance loss. Simulation examples and several case studies in various applications illustrate the effectiveness of the proposed framework.

*Key words*: anomaly detection; compressive sensing; image processing; smooth sparse decomposition.

## 1 Introduction

High-quality image sensing systems are widely used in manufacturing processes for product quality monitoring and fault diagnosis. The resolution and acquisition rate of such systems increase significantly benefiting from the rapid development of image sensing technology. For example, in a hot rolling process, an in-situ image-based sensor can detect a micrometer size seam on a rolling bar at a speed of up to 225 miles per hour (Yan et al., 2018). For another example, to monitor solar activity, satellites can capture high-resolution solar images with a high acquisition rate, producing terabytes of data per day (Wang et al., 2018). To achieve real-time inspection, a large volume of high-resolution images needs to be transmitted and processed in real-time. Such a large volume of high-resolution image data poses a big challenge not only on the speed of image processing algorithms but also the storage and transmission of the data itself.

Matrix decomposition-based image processing techniques are widely used in image-based process monitoring and anomaly detection. It achieves this goal by integrating the prior for background and anomaly components into the optimization problem. In terms of utilizing the low-rank and sparse property, Robust Principal Component Analysis (RPCA) was first proposed by Candès et al. (2011) to decompose a data matrix into a low-rank and an element-wise sparse component. One of its famous applications is the dynamic foreground and static background separation (Bouwmans & Zahzah, 2014). Following this approach, numerous algorithm variants have been proposed including Outlier Pursuit (Xu et al., 2012) which aims to decompose the data matrix into a low-rank component and column-wise sparse component, low-rank plus compressed sparse decomposition (Mardani et al., 2013) which aims to decompose the data matrix into a low-rank and a compressed sparse component and so on. For utilizing smooth and sparse properties, Smooth and Sparse Decomposition (SSD) methods (Minaee et al., 2015; Yan et al., 2017) are proposed for anomaly detection in images with smooth backgrounds. Following this approach, several explorations have been conducted including Spatio-temporal smooth sparse decomposition (ST-SSD) (Yan et al., 2018), Additive tensor decomposition (ATD) (Mou et al., 2021), and so on. By adopting the matrix decomposition approach, both the background and anomaly can be captured without detection time delay. However, due to the requirement of storage, transmission, and processing of the whole image signal, it cannot be applied in the scenario with low transmission bandwidth but high processing speed requirements, for example, the solar flare detection application (Augusto et al., 2011).

---

[*] No data ethics considerations are foreseen related to this paper.

[†] Dr. Shi is the corresponding author.

To mitigate the data storage, transmission burden and improve sensing efficiency, Compressive Sensing (CS) (Candes et al., 2006) has been proposed, in which the data is directly collected in a compressed form and then reconstructed accurately with high probability. More specifically, suppose that the original signal is a *sparse* vector $y \in \mathbb{R}^n$, the main idea is to store and transmit a small set of compressive measurements $y' = Ay \in \mathbb{R}^p$, where $A \in \mathbb{R}^{p \times n}$ is an underdetermined sensing matrix ($p \ll n$) satisfying specific properties. Then, the original signal can be reconstructed from its compressed form $y'$, on which assorted image processing algorithms can be applied for defect detection and so on. For a comprehensive review of CS, please refer to Marques et al. (2018) and Rani et al. (2018). Even though promising, the naïve approach which tries to first reconstruct the image from the compressed measurement and then apply matrix decomposition algorithms for anomaly detection has two issues:

(i) For smooth plus sparse signals, the existence of such a sensing matrix $A$ satisfying specific properties is unknown.

(ii) The reconstruction process is usually computationally intensive (Marques et al., 2018), which on the other hand, slows down the overall computational speed of the defect detection algorithms.

Recently, to integrate the CS with matrix decomposition algorithms, Waters et al. (2011) proposed an SpaRCS method to recover Low-Rank and sparse matrices directly from compressive measurements, and Tanner and Vary (2020) gave a rigorous performance discussion. However, those methods do not consider the smooth plus sparse decomposition problems and are not efficient in dealing with high-order data.

In this paper, we discuss the possibility of adopting compressive data acquisition systems for image-based quality monitoring and fault detection in applications where the background is smooth, and defects are sparse. To achieve so, we propose a Compressed Smooth Sparse Decomposition (CSSD) framework. In this framework, the signal processing algorithms are directly applied to the compressed data and no reconstruction step is needed. By doing so, a significant cost reduction in sensing, storage, and transmission as well as a boost in the speed of image processing algorithms, can be achieved with negligible performance loss. We also established the theoretical foundation of adopting such a compressive data acquisition system for smooth plus sparse signals as well as the performance guarantee of the proposed algorithm. To further improve its performance in high-order scenarios, a Kronecker Compressed Smooth Sparse Sensing (KronCSSD) is proposed.

The remainder of this paper is organized as follows. In Section 2, we present the CSSD framework. In Section 3, we use simulation studies to validate the proposed framework. In Section 4, we demonstrate the proposed framework using several case studies. Finally, Section 5 concludes the paper.

## 2 Compressed Smooth Sparse Decomposition Framework

In this section, we will present the proposed CSSD framework. As mentioned in the Introduction, we aim to design a fast and data-efficient method for sparse anomaly (sparse signal component) detection in signals with a smooth background (smooth signal component). For simplicity, we first discuss the methodology for one-dimensional (1-D) signals and then generalize it to n-D images. Figure 1 provides an overview of the proposed methodology. There are three stages in the proposed methodology: (i) the signal is acquired in its compressed form through compressive measurement; (ii) Then the compressed data is transmitted to the server; (iii) Finally, a decomposition algorithm will be applied directly to the compressed signal to decompose it into its corresponding smooth and sparse signal components.

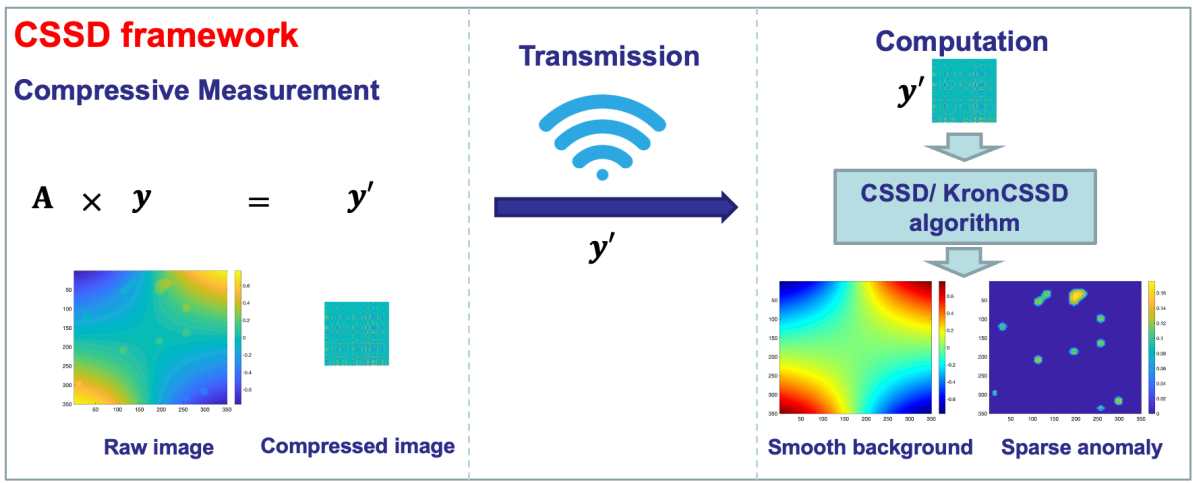

Figure 1. An overview of the CSSD/KronCSSD framework.

Mathematically, let $y \in \mathbb{R}^n$ be a smooth plus sparse signal (which will be defined formally in Section 2.1), we aim to store and transmit a small set of compressive measurements $y'$, i.e., $y' = Ay$, then reconstruct the smooth and sparse signal components from $y'$. To achieve so, there are three questions to be addressed:

(i) What is a smooth plus sparse signal?

(ii) How to compress such a signal? and

(iii) How to reconstruct such a signal from compressive measurements?

The remainder of this section is organized as follows to answer those three questions. We start with a formal definition of 1-D smooth plus sparse signal in Section 2.1. Based on that, we introduce the compressive measurement method and discuss its theoretical properties for such signals in Section 2.2. In section 2.3, we present the proposed CSSD framework that can directly reconstruct the smooth and sparse signal components from the compressive measurement by solving the following optimization problem:

$$\min_{\theta, \theta_a} \|\theta_a\|_1$$

s.t. (1)

$$\|A(B\theta + B_a \theta_a) - y'\|_2 \le \epsilon_1$$

where $B$ and $B_a$ are bases, $\theta$ and $\theta_a$ are corresponding coefficients, and $\epsilon_1$ is the bound for measurement error. The reconstruction accuracy is also characterized theoretically. Then, we generalize the CSSD algorithm to n-D using Kronecker Compressive Sensing (KCS) (Duarte & Baraniuk, 2011) and proposed a KronCSSD formulation in Section 2.4. Finally, in section 2.5, we present the advantage of the proposed framework and give the strategy of selecting compressive ratio, tuning parameters and bases in practice.

## 2.1 The set of smooth plus sparse signals

In this section, we define the set of smooth plus sparse signals mathematically. The smooth signals originate from the spline smoothing (De Boor & De Boor, 1978; Eilers & Marx, 1996) where the raw signal is approximated by a linear combination of a set of spline basis functions for smooth interpolation and denoising, and a spline regression technique is usually utilized. To improve the regression robustness with respect to outliers, outliers are explicitly accounted for in the regression model as a sparse component of the raw signal (Giannakis et al., 2011; Mateos & Giannakis, 2011). This idea was further generalized to incorporate the special structure of outliers (Yan et al., 2017, 2018).

As a summary, a 1-D signal $y \in \mathbb{R}^n$ is defined as a smooth plus sparse signal if it can be decomposed into two signal components: (i) a smooth signal $m \in \mathbb{R}^n$ in a low dimensional subspace spanned by a set of smooth bases, i.e., $m = B\theta$, where $B \in \mathbb{R}^{n \times r}$ is a basis matrix with $r \ll n$; (ii) a sparse signal $a \in \mathbb{R}^n$ in a relatively high dimensional subspace spanned by a set of predefined bases of which the coefficients admit sparse property, i.e., $a = B_a \theta_a$ where $B_a \in \mathbb{R}^{n \times q}$ is a basis matrix with $q \le n$ and $\boldsymbol{\theta}_a \in \mathbb{R}^q$ is an $s$-sparse vector, i.e., $\|\boldsymbol{\theta}_a\|_0 \leq s$. Given a smooth plus sparse signal $\boldsymbol{y}$, we define the aforementioned decomposition as smooth sparse decomposition (SSD), i.e., $\boldsymbol{y} = \boldsymbol{m} + \boldsymbol{a}$.

To ensure the uniqueness of SSD in a nontrivial case when $n \leq r + q$, the following definition is introduced.

**Definition 2.1.** *Local support property of $\boldsymbol{B}_a$, $\boldsymbol{B}_a \in \mathbb{R}^{n \times q}$. $\boldsymbol{B}_a$ only has local support such that each column of $\boldsymbol{B}_a$ only has non-zero values inside a specific interval. The length of this interval is defined as $l(\boldsymbol{B}_a)$.* ∎

Notice that $l(\boldsymbol{B}_a) \in \{1, \ldots, n\}$. The local support property with a small $l$ ensures the sparsity of $\boldsymbol{a}$. For example, the B-spline basis has a local support property (Unser, 1999).

**Definition 2.2.** *Incoherence condition of $\boldsymbol{B}$, $\boldsymbol{B} \in \mathbb{R}^{n \times r}$. Let $\boldsymbol{B} = \boldsymbol{U}\boldsymbol{\Sigma}\boldsymbol{V}^T$ be the reduced singular value decomposition (SVD) of $\boldsymbol{B}$, where $\boldsymbol{U} \in \mathbb{R}^{n \times r}, \boldsymbol{\Sigma} \in \mathbb{R}^{r \times r}$ and $\boldsymbol{V} \in \mathbb{R}^{r \times r}$. Its incoherence condition parameter $\mu(\boldsymbol{B})$ is defined as the smallest value such that*

$$\max_{i \in \{1, \ldots, r\}} \|\boldsymbol{U}^T \boldsymbol{e}_i\|_2 \leq \sqrt{\frac{\mu(\boldsymbol{B})r}{n}},$$

*where $\boldsymbol{e}_i$ is the $i$-th standard basis vector in $\mathbb{R}^n$.* ∎

Notice that $\mu(B) \in [1, \sqrt{n/r}]$. The incoherent condition with a small $\mu$ ensures that the $\boldsymbol{m}$ is not sparse (Candès et al., 2011).

The following theorem ensures the uniqueness of the SSD decomposition.

**Theorem 2.1.** *If $\mu(\boldsymbol{B}) < n(2rsl)^{-1}$, then the SSD decomposition is unique with respect to $\boldsymbol{m}$ and $\boldsymbol{a}$.* ∎

Theorem 2.1 gives the conditions that the smooth plus sparse signal can be uniquely decomposed into a smooth part and sparse part. The proof of Theorem 2.1 is in Appendix A.

Formally, we define the set of smooth plus sparse signals as follows.

**Definition 2.3.** *The set of smooth and sparse signals is defined as $MS_{r,s,\mu,l}$:*

$MS_{r,s,\mu,l} = \{\boldsymbol{y} \in \mathbb{R}^n \mid \boldsymbol{y} = \boldsymbol{B}\boldsymbol{\theta} + \boldsymbol{B}_a\boldsymbol{\theta}_a, \boldsymbol{B}_a \in \mathbb{R}^{n \times q}, l(\boldsymbol{B}_a) = l, \boldsymbol{\theta}_a \in \mathbb{R}^q, \|\boldsymbol{\theta}_a\|_0 \leq s, \boldsymbol{B} \in \mathbb{R}^{n \times r},$
$\mu(\boldsymbol{B}) = \mu < n(2rsl)^{-1}, \boldsymbol{\theta} \in \mathbb{R}^r\}.$ ∎

For such a smooth plus sparse signal, how to compress it while ensuring the reconstruction performance will be discussed in the next section.

## 2.2 Compressive sensing for smooth plus sparse signals

As mentioned in the Introduction, to ensure the reconstruction performance, the sensing matrix $\boldsymbol{A}$ has to satisfy the so-called Restricted Isometry Property (RIP) (Candes, 2008). It has been proved that a random matrix can satisfy the RIP property for the sparse signal (Candes, 2008), low-rank signal (Recht et al., 2010), and the rank plus sparse signal (Tanner & Vary, 2020). However, the existence of such a matrix for the smooth plus sparse signal, which is the foundation of adopting compressed data acquisition techniques in applications with smooth background and sparse anomalies, is still unknown.

In this section, we will discuss the existence of such a sensing matrix. Before stating the result, we will first present the relevant definitions that are necessary to derive the main result.

**Definition 2.4.** *RIP for $MS_{r,s,\mu,l}$. Let $A \in \mathbb{R}^{p \times n}$ be a linear measurement matrix. For every quadruple $(r, s, \mu, l)$, define the restricted isometry constant (RIC) $\delta_{r,s,\mu,l}$ to be the smallest positive constant such that*

$$(1 - \delta_{r,s,\mu,l})\|y\|_2 \leq \|Ay\|_2 \leq (1 + \delta_{r,s,\mu,l})\|y\|_2, \quad \forall\, y \in MS_{r,s,\mu,l}.$$

If such a $\delta_{r,s,\mu,l} \in (0,1)$ exists, we say that $A$ satisfies the RIP. ∎

**Theorem 2.2.** *Suppose that $\delta_{r,2s,\mu,l} < 1$ for some integer $r, s, l \geq 1$ and positive numbers $\mu < n(2rsl)^{-1}$, then, there is a $y_0$ in the set $MS_{r,s,\mu,l}$, which is the only solution for $Ay_0 = b$.* ∎

Theorem 2.2 guarantees the uniqueness of smooth plus sparse signal that satisfies the sensing equation when $A$ satisfies the RIP. The proof of Theorem 2.2 is in Appendix B.

Next, we prove that for the set of smooth plus sparse signals, $MS_{r,s,\mu,l}$, there exists such a matrix $A$ satisfying the RIP property with RIC=$\delta_{r,s,\mu,l}$ with high probability.

Notice that the RIP for a matrix is difficult to verify. A suitable set of random matrices that obey the RIP for the set of sparse vectors with high probability (Recht et al., 2010; Tanner & Vary, 2020) is defined as follows:

**Definition 2.5.** *Nearly isometric matrices (Baraniuk et al., 2008). Let $A \in \mathbb{R}^{p \times n}$ be a random variable that takes values in linear maps from $\mathbb{R}^n$ to $\mathbb{R}^p$, then for any $y \in \mathbb{R}^n$, $A$ is nearly isometrically distributed if*

(i) $\mathbb{E}[\|Ay\|_2^2] = \|y\|_2^2$,

(ii) $Pr(|\|Ay\|_2^2 - \|y\|_2^2| \geq \epsilon\|y\|_2^2) \leq 2e^{-pc_0(\epsilon)},\ 0 < \epsilon < 1,$

where $c_0(\epsilon)$ is a constant that only depends on $\epsilon$. ∎

The $p \times n$ matrix with independent, identically distributed (i.i.d.) Gaussian entries satisfies those two properties (Baraniuk et al., 2008), i.e., $A_{ij} \sim \mathcal{N}\left(0, \frac{1}{p}\right)$, with $c_0(\epsilon) = \epsilon^2/4 - \epsilon^3/6$. There are also other distributions satisfying the nearly isometric property, such as the $p \times n$ matrix with i.i.d. Bernoulli entries and their related distribution (Baraniuk et al., 2008).

Then, the following theorem states that the nearly isometric matrices can also serve as the sensing matrix for smooth plus sparse signals and gives the magnitude of the number of linear measurements.

**Theorem 2.3.** *Let $A \in \mathbb{R}^{p \times n}$ be a matrix from the families described in Definition 2.5. Further, assume $\mu < n(2rsl)^{-1}$ and basis matrix $B_a$ for sparse signal component satisfies the RIP with RIC $\delta_{B_a,s} \in (0,1)$, i.e., $\delta_{B_a,s}$ to be the smallest positive constant such that*

$$(1 - \delta_{B_a,s})\|\theta_a\|_2 \leq \|B_a \theta_a\|_2 \leq (1 + \delta_{B_a,s})\|\theta_a\|_2, \quad \forall\, \theta_a \in \{\theta_a \in \mathbb{R}^q, \|\theta_a\|_0 \leq s\}.$$

*For a given $\delta \in (0,1)$, there exists constants $c_1, c_2 > 0$ depending only on $\delta$, such that the RIC for $MS_{r,s,\mu,l}$ is upper bounded by $\delta$, with the probability of at least $1 - exp(-c_1 p)$, whenever*

$$p \geq c_2 \left( \ln 2 + r \ln \frac{24}{\delta} \tau_1 + s \left( 1 + \ln \frac{24}{\delta} \tau_0 + \ln \frac{n}{s} \right) \right),$$

where $\eta = \sqrt{\frac{\mu r s l}{n}}$, $\tau_0 = \frac{1}{\sqrt{(1-\delta_{B_a,s})(1-\eta^2)}}$, $\tau_1 = \|B^\dagger\|_2 \left( 1 + \frac{1}{\sqrt{1-\eta^2}} \right)$, and $B^\dagger = (B^T B)^{-1} B^T$. ∎

Theorem 2.3 states that, if the nearly isometric matrix is selected as the sensing matrix, the RIC for $MS_{r,s,\mu,l}$ is upper bounded with high probability. In practice, $B$ and $B_a$ are pre-specified based on the understanding/ engineering knowledge of the process (please refer to Section 2.5.4 for more detail). Theorem 2.3 also provides a guidance on the magnitude of linear measurements $p$ which determines the compressive measurement matrix $A$.

The proof of Theorem 2.3 is in Appendix C.

In this section, we answered two fundamental questions: (i) How to compress the smooth plus sparse signal (design the compressive measurement matrix $A$); (ii) How many linear measurements are needed to preserve the information in smooth plus sparse signals with high probability.

In the next section, we will discuss the problem formulation to recover the smooth and sparse signal components simultaneously from the compressed signal using the CSSD framework.

## 2.3 Compressed smooth sparse decomposition

As mentioned in the Introduction, one way to recover the smooth component and sparse component is first reconstructing the compressed image and then using the SSD algorithm. However, this will slow down the speed of the defect detection algorithm. Instead, we propose to solve the one-step convex relaxation problem (1). One natural question is that can we recover the smooth and sparse signals from the compressed measurement $y'$ by solving Problem (1) and what is the accuracy. The following theorem guarantees the recovery performance of the proposed convex relaxation Problem (1).

**Theorem 2.4.** *Let $A \in \mathbb{R}^{p \times n}$ be a matrix from the families described in Definition 2.5. Let the signal $y = m_0 + a_0 \in MS_{r,s,\mu,l}$, where $m_0 = B\theta_0$ and $a_0 = B_a \theta_{a0}$. Assume that Problem (1) is feasible and let the optimal solution to be $m^* = B\theta^*$ and $a^* = B_a \theta_a^*$. Assume the basis matrix $B_a$ for sparse signal component satisfies the RIP with RIC $\delta_{B_a,2s} \in (0,1)$. Let $a = (1 + \alpha_1 + \alpha_2)\gamma^2 + 2\alpha_2 + 2$ and $c = 1 - \gamma^2 \alpha_1 \alpha_2 - \alpha_2^2$, where $\alpha_1 = \frac{\eta}{1-\eta^2}$, $\alpha_2 = \frac{\sqrt{2}\eta}{1-2\eta^2}$, $\gamma = \sqrt{\frac{1+\delta_{B_a,2s}}{1-\delta_{B_a,2s}}}$, $\eta = \sqrt{\frac{\mu r s l}{n}}$. Suppose that $r, s, l \in \mathbb{N}$ and $\mu < n(2rsl)^{-1}$, such that $c > 0$ and the $\delta_{r,3s,\mu,l} \in (0, c/a)$, then*

$$\|a_0 - a^*\|_2 = \|B_a \theta_a^0 - B_a \theta_a^*\|_2 \leq C_a \epsilon_1,$$

and

$$\|m_0 - m^*\|_2 = \|B\theta_0 - B\theta^*\|_2 \leq C_m \epsilon_1,$$

where

$$C_a = \frac{(1 + \gamma^2)(1 + \alpha_2)\sqrt{1 + \delta_{r,3s,\mu,l}}}{c - a\delta_{r,3s,\mu,l}},$$

and

$$C_m = \frac{\sqrt{1+\delta_{r,3s,\mu,l}} + \left(\delta_{r,3s,\mu,l} + \frac{\gamma^2}{1+\gamma^2}\alpha_1 + \frac{1}{1+\gamma^2}\alpha_2\right)C_a}{\left(1-\delta_{r,3s,\mu,l}\right)}. \qquad \blacksquare$$

Theorem 2.4 gives the conditions that the proposed convex relaxation problem (1) can recover the true smooth and sparse signal up to a constant times the noise bound.

The proof of Theorem 2.4 is in Appendix D.

Notice that one advantage of the proposed CSSD framework is that it is compatible with existing decomposition algorithms. For example, for the SSD algorithm proposed by Yan et al. (2017), the problem formulation becomes:

$$\min_{\theta,\theta_a} \|y' - A(B\theta + B_a\theta_a)\|_2^2 + \lambda\|\theta_a\|_1, \qquad (2)$$

which can be solved efficiently by using the algorithm proposed by Yan et al. (2017).

## 2.4 Kronecker compressed smooth sparse decomposition

In the previous section, we discussed the CSSD framework for 1-D signal. In this section, we will generalize the proposed CSSD to KronCSSD framework for high order tensor data.

Let $\mathcal{Y} \in \mathbb{R}^{n_1 \times \ldots \times n_d}$ be the original signal and $y \in \mathbb{R}^N$ be its corresponding vectorized signal, i.e., $y = \text{vec}(\mathcal{Y})$ and $N = \prod_{i=1}^d n_i$. $A \in \mathbb{R}^{p \times N}$ be a measurement matrix satisfying the RIP. Let $y' \in \mathbb{R}^p$ be the compressed data, such that $y' = Ay$. $B = \otimes_{i=1}^d B_i$, $B_a = \otimes_{i=1}^d B_{ai}$ are the known bases for smooth and sparse components, respectively. Notice that Problem formulation (1) can still be used for recovering the high order smooth and sparse signal components from the compressive measurement. However, there are two issues: (i) in practice, the global CS measurements matrix $A$ is hard to realize using CS device (Duarte & Baraniuk, 2011); (ii) the resulting bases $B$ and $B_a$ can be extremely large as the dimension of the data increases which will not only cause a big challenge in the storage of such a large matrix but also result in the computational issue in handling such large matrices.

In reality, the high dimensional tensor data usually have low-rank properties along each mode, which has been extensively exploited in tensor low-rank modeling techniques such as CP/Tucker decompositions (Kolda & Bader, 2009). This makes it possible for designing a sensing matrix for each mode, which is called Kronecker CS (Duarte & Baraniuk, 2011). Inspired by the KCS method, we propose the KronCSSD framework formulation as follows:

Let $A_i \in \mathbb{R}^{p_i \times n_i}$ be a measurement matrix with RIP for each mode of the tensor, then we have the following formulation,

$$\min_{\Theta,\Theta_a} \|\text{vec}(\Theta_a)\|_1$$

s.t. $\qquad\qquad\qquad\qquad\qquad\qquad\qquad\qquad\qquad\qquad\qquad\qquad\qquad\qquad (3)$

$$\|\text{vec}(\mathcal{Y}_1 - \mathcal{Y}')\|_2 \leq \epsilon_1,$$

$$\mathcal{Y}_1 = \Theta \times_1 (A_1 B_1) \times_2 \ldots \times_d (A_d B_d) + \Theta_a \times_1 (A_1 B_{a1}) \times_2 \ldots \times_d (A_d B_{ad}),$$

where $\mathcal{Y}' = \mathcal{Y} \times_1 A_1 \times_2 \ldots \times_d A_d$ is the compressive measurement. $\Theta \in \mathbb{R}^{r_1 \times \ldots \times r_d}$ and $\Theta_a \in \mathbb{R}^{q_1 \times \ldots \times q_d}$ are the basis coefficients for the smooth and sparse signal components, respectively.

The proposed KronCSSD framework is a non-trivial generalization of the CSSD framework for 1-D signals. Its performance will be shown empirically in simulation and case studies. The theoretical discussion of the KronSSD framework is left for future work.

For example, for a 2-D image $Y \in \mathbb{R}^{n_1 \times n_2}$, when adopting the SSD algorithm (Yan et al., 2017), the problem formulation becomes:

$$\min_{\Theta,\Theta_a} \left\| Y' - A_1 B_1 \Theta B_2^T A_2^T - A_1 B_{a1} \Theta_a B_{a2}^T A_2^T \right\|_2^2 + \lambda \|\text{vec}(\Theta_a)\|_1, \tag{4}$$

where $Y'$ is the compressed image, i.e., $Y' = A_1 Y A_2^T$. It can be solved efficiently by using the algorithm proposed by Yan et al. (2017).

### 2.5 Discussion

#### 2.5.1 Advantages of the proposed CSSD/ KronCSSD methods

In this section, we will give a brief discussion about the advantages of the proposed CSSD/ KronCSSD methods. We define the compressive ratio as $c = \prod_{i=1}^{d} p_i / n_i$. The smaller the compressive ratio, the fewer data will be transmitted. The proposed CSSD/ KronCSSD methods have the following characteristics:

(i) We propose to directly acquire the compressed image, which not only reduces the sensing cost but also reduces the data transmission and storage cost by $\prod_{i=1}^{d} p_i / n_i$ times.

(ii) The smooth and sparse signal components can be recovered by solving a smaller scale convex optimization problem with input data $\prod_{i=1}^{d} p_i / n_i$ times smaller than that of the original problem, which significantly boosts the computation.

#### 2.5.2 Compressive ratio selection in practice

Theorem 2.3 and Theorem 2.4 show that the 1-D smooth plus sparse signal can be recovered with high probability from compressive measurement if the compressive ratio is above a specific threshold. However, there are several parameters (such as $c_1, c_2$) that are difficult to obtain when calculating the threshold in practice. Below, we propose a practical procedure of selecting the compressive ratio utilizing the historical data. Suppose some training signals with their real background and anomalies are available, the compressive ratio can be selected with the following guidelines:

(i) If there is a requirement for reconstruction accuracy for smooth and sparse signal components, then $\hat{p}$ is chosen as the smallest value that satisfies such requirement.

(ii) If there is no such requirement, we recommend choosing the compressive ratio corresponding to the sharp change point of the slope of the loss function-compressive ratio curve. This point exists because of the existence of such a threshold after which the reconstruction with high probability is guaranteed by Theorem 2.3 and Theorem 2.4. We will demonstrate this in the simulation study in Section 3.1.

For high-order tensor data, the selection of the $p_i$ for each mode can be challenging. We provide some empirical guidelines as follows:

(i) If the smoothness of the background is similar along different modes, a unified compressive ratio is recommended.

(ii) If the smoothness of the background is different, more sensing budget should be allocated to the mode along which the background is less smooth.

(iii) We recommend fixing the ratio among $p_i$ along different modes and adopting steps (i) and (ii) for 1-D signal to determine the compressive ratio.

### 2.5.3 Tuning parameter selection in practice

Notice that in problem formulations (1) and (3), there is a hyperparameter $\epsilon_1$ indicating the bound for measurement noise. If the measurement error bound is known from the accuracy of the measurement device, it can be directly used here. Otherwise, a cross-validation step using historical data is recommended. Similarly, there is a tuning parameter $\lambda$ in their corresponding Lagrangian form Eqs (2) and (4), which controls the sparsity of the decomposed anomaly. Cross-validation can be used to determine this parameter. For more detail, please refer to Yan et al. (2017).

### 2.5.4 Bases selection in practice

The bases $\boldsymbol{B}$ and $\boldsymbol{B_a}$ are pre-specified based on the understanding/ engineering knowledge of the process. The selection of such bases is discussed in detail in Section 3.4 of Yan et al. (2017). In general, any smooth basis, such as splines or kernels, can be used for the background. For sparse anomalies such as small regions scattered over the background or in the form of thin lines, an identity basis is recommended. Linear (quadratic) B-splines are recommended for anomalous regions with sharp corners (curved boundaries). However, to ensure the uniqueness of the smooth sparse decomposition, we do require the bases $\boldsymbol{B}$ and $\boldsymbol{B_a}$ to satisfy specific properties, such as those mentioned in Definition 2.1, Definition 2.2 and Theorem 2.1, which should be checked for selected bases. We will demonstrate this in the simulation study in Section 3.1.

## 3 Simulation Study

In this section, we will demonstrate the proposed CSSD and KronCSSD framework with simulation studies. First, the CSSD method is applied on 1-D signals in Section 3.1 and then we apply the KronCSSD method on 2-D images in Section 3.2.

### 3.1 CSSD on 1-D signal

A 1-D signal ($\boldsymbol{y} \in \mathbb{R}^n$) is assumed to be a superposition of smooth signal component ($\boldsymbol{m} = \boldsymbol{B\theta}$), sparse signal component ($\boldsymbol{a} = \boldsymbol{B_a\theta_a}$), and noise $\boldsymbol{e}$, i.e., $\boldsymbol{y} = \boldsymbol{m} + \boldsymbol{a} + \boldsymbol{e}$. To demonstrate the proposed CSSD framework, we will first conduct compressive data acquisition on the raw signal, i.e., $y' = Ay$. Then, the data reconstruction and decomposition are achieved in one step. The performance is evaluated by the relative error between the true signal $a$ and the reconstructed one $\hat{a}$, i.e., $\|a - \hat{a}\|_2 / \|a\|_2$, and $\|m - \hat{m}\|_2 / \|m\|_2$ for the smooth background.

In the simulation study, we generate a 1-D signal from $MS_{r,s,\mu,l}$ and let $n = 1000$. The smooth background is generated from a random linear combination of B-Spline bases with 3 knots ($r = 10$), i.e., $m = B\theta$ where $B \in \mathbb{R}^{n \times r}$ and $\theta \in \mathbb{R}^r$ is a random vector such that $\theta_i \sim N(0,1)$, $i \in \{1, \dots, 4\}$. The incoherence condition parameter $\mu = \mu(B) = 0.82$. The sparse signal component is generated by a sparse random linear combination of degree 2 B-Spline bases with 500 knots ($q = 500, l = 4$), i.e., $a = B_a \theta_a$ where $B_a \in \mathbb{R}^{n \times q}$ and $\theta_a \in \mathbb{R}^q$ is a 4-sparse random vector ($s = 4$) such that its non-zero elements follow i.i.d. standard normal distribution. Notice that the RIC, $\delta_{B_a,s}$ for matrix $B_a$ is hard to calculate in general. However, there is a loose upper bound that can be used, which is $\delta_{B_a,s} \le \delta_{B_a,p} = \max\{\lambda_{max} - 1, 1 - \lambda_{min}\} = 0.60$ where $\lambda_{max}$ and $\lambda_{min}$ are the maximum and minimum singular values of $B_a$. The noise signal is generated as a random vector $e \in \mathbb{R}^n$ is a random vector such that $e_i \sim N(0, 0.001^2)$, $i \in \{1, \dots, n\}$. Figure 2 shows a sample raw signal and its smooth, sparse components.

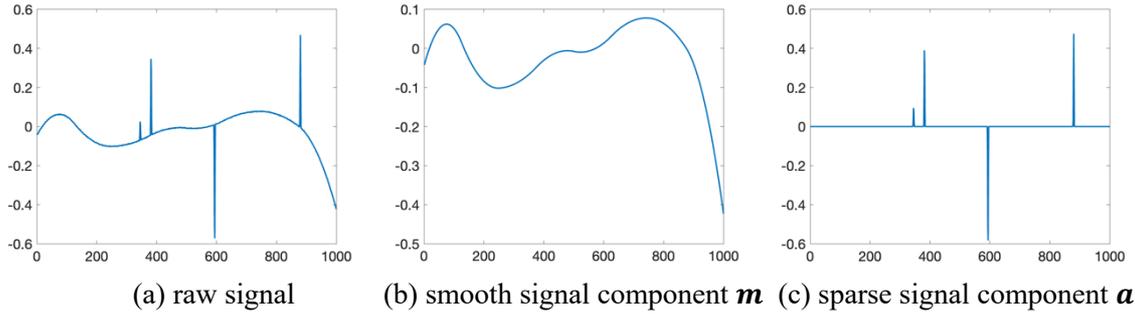

(a) raw signal  (b) smooth signal component $m$  (c) sparse signal component $a$

Figure 2. A sample raw signal and its smooth, sparse component

Before we state the result, we first check the assumption of Theorem 2.3 and Theorem 2.4. For Theorem 2.3, it is easy to check that $0.82 = \mu \le \frac{n}{2rsl} = \frac{1000}{2 \times 10 \times 4 \times 4} = 6.25$, and also $\delta_{B_a,s} \le 0.60 \in (0,1)$. For Theorem 2.4, $\delta_{B_a,2s} \le 0.60 \in (0,1)$, $c = 0.37 > 0$ and $\delta_{r,3s,\mu,l} \in (0,0.04)$. Notice that the range for $\delta_{r,3s,\mu,l}$ is small in this case because of the loose upper bound for $\delta_{B_a,2s}$, which can be further improved. This indicates that the smooth and sparse signal can be recovered by the proposed algorithm with high probability provided that the compressive ratio being above a specific threshold, which is demonstrated by the following observation:

The simulation is repeated 100 times and the average log relative error for the background and sparse signal components with respect to compressive ratio are shown in Figure 3 (a) and (b), respectively. We can observe a large error at the beginning, and it drops very fast with the increase of compressive ratio. Then, above a threshold (0.1 approximately) of compressive ratio, the error becomes small (below 3%) and the decrease of error becomes less significant, which demonstrates the effectiveness of the reconstruction algorithm. The threshold of 0.1 can be chosen as the compressive ratio mentioned in Section 2.5.2.

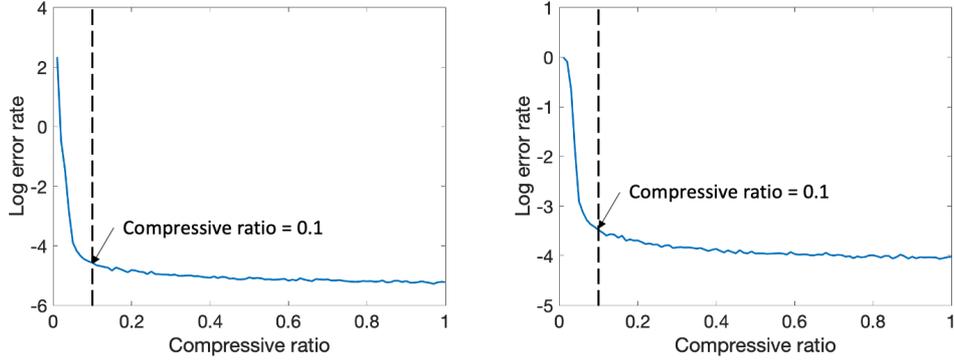

(a) Log relative error for smooth component   (b) Log relative error for sparse component

Figure 3. Average log relative reconstruction error

The reconstructed signal components in one of the 100 simulations are shown in Figure 4. We can observe that when adopting the compressive ratio of 0.1, both the smooth and sparse signal components can be reconstructed with high accuracy.

We also vary the magnitude of sparse signals to examine the reconstruction performance of the proposed method. The result and analysis are provided in Appendix H.

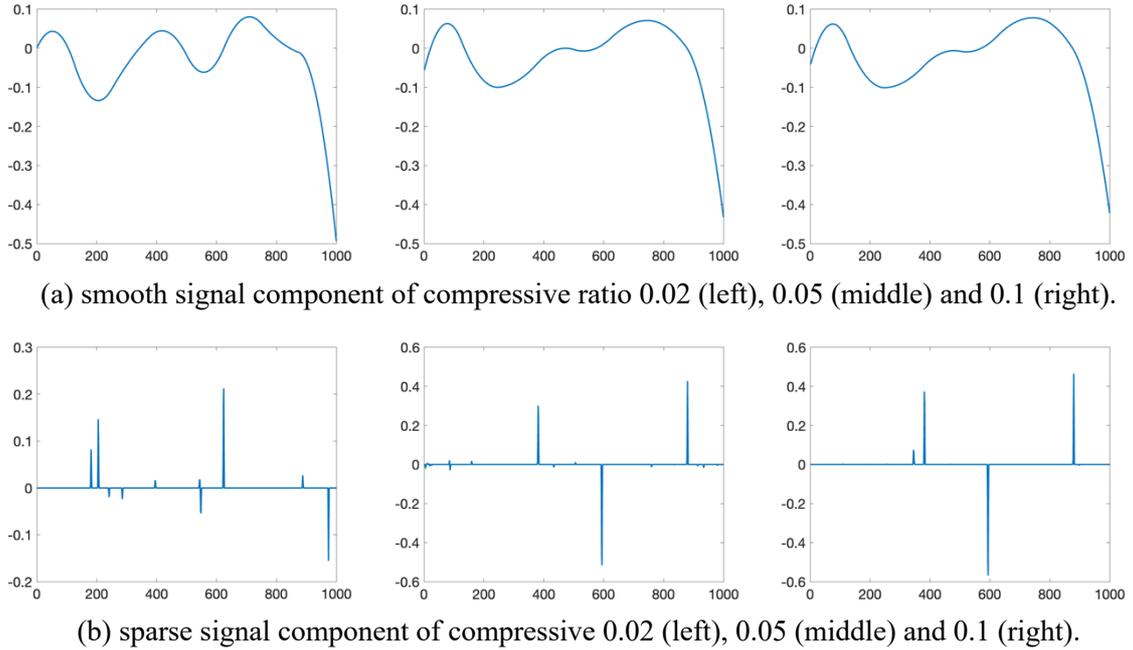

(a) smooth signal component of compressive ratio 0.02 (left), 0.05 (middle) and 0.1 (right).

(b) sparse signal component of compressive 0.02 (left), 0.05 (middle) and 0.1 (right).

Figure 4. Reconstructed smooth and sparse signal components

### 3.2 CSSD on 2-D image

In this simulation study, we aim to decompose an image into the smooth background, sparse anomalies, and noise. A $350 \times 350$ image with smooth background and the sparse anomaly is generated similar to Yan et al. (2017), i.e., $Y = M + A + E$, where $M$ is the smooth background, $A$ is the sparse anomalies and $E$ is i.i.d. Gaussian noise such that $E_i \sim NID(0, \sigma^2)$. The smooth background is generated from a linear combination of B-Spline bases with 3 × 3 knots and the anomalies are generated from a sparse linear combination of B-Spline bases with 88 × 88 knots. The background and anomalies are shown in Figure 5.

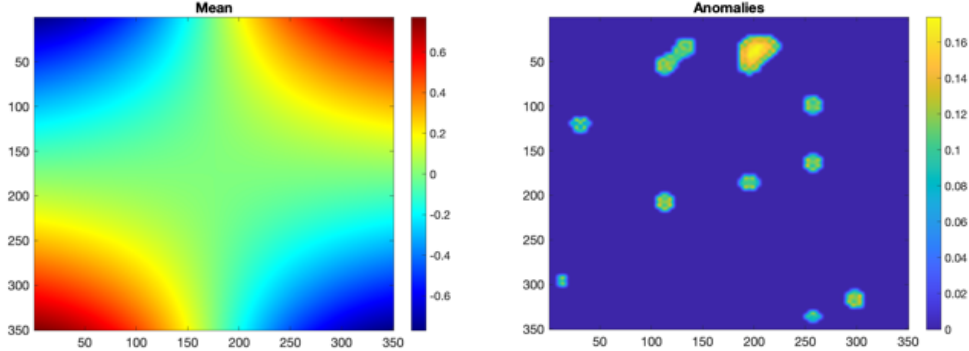

Figure 5. True background and anomalies

We can see that the anomaly size covers a large range. The mean absolute value of the background plus anomaly is $\mu = 0.21$. In this simulation, we first study the reconstruction performance under a noise-free scenario. Then, we increase the noise level to test the robustness of the algorithm.

*3.2.1 Noise-free case*

In this section, we vary the compressive ratio from 4% to 100%. For each compressive ratio, we simulate 100 times and the average False negative rate (FNR) and False positive rate (FPR) are reported in Figure 6. The FPR is defined as the portion of normal pixels predicted as anomaly, i.e.,

$$FPR = \frac{\sum_{i,j}(1 - I_{A\neq 0}\{A(i,j)\})I_{\widehat{A}\neq 0}\{\widehat{A}(i,j)\}}{\sum_{i,j}(1 - I_{A\neq 0}\{A(i,j)\})},$$

and the FNR is defined as the portion of anomalous pixels predicted as normal background, i.e.,

$$FNR = \frac{\sum_{i,j} I_{A\neq 0}\{A(i,j)\}(1 - I_{\widehat{A}\neq 0}\{\widehat{A}(i,j)\})}{\sum_{i,j} I_{A\neq 0}\{A(i,j)\}},$$

where $A$ is the true anomaly, $\widehat{A}$ is the predicted anomaly and $I_\Omega(\cdot)$ is the indicator function, i.e.,

$$I_\Omega(x) = \begin{cases} 1, if\ x \in \Omega \\ 0, otherwise \end{cases}.$$

From Figure 6, we can see that both the FPR and FNR ratios decrease as the compressive ratio increases. The FPR drops so fast that when the compressive ratio achieves 8%, there is no false alarm, which is desired for the anomaly detection algorithm. The FNR drops slower, and less than 5% of the anomaly pixels are ignored when the compressive ratio achieves 8%. However, it does not miss any cluster of the anomalies even though some of them are small.

The recovered sparse signal components are shown in Figure 7 for compressive ratio 4% (Figure 7 (a)), 8% (Figure 7 (b)), 33% (Figure 7 (c)), and 73% (Figure 7 (d)). For comparison, we apply the SSD algorithm (Yan et al., 2017) of which the FNR and FPR are 0, and computation time 0.16s.

We record the computation time in Figure 8. A significant boosting of the computation is observed when the compressive ratio is 8%, which speeds up the SSD algorithm by 4.3 times.

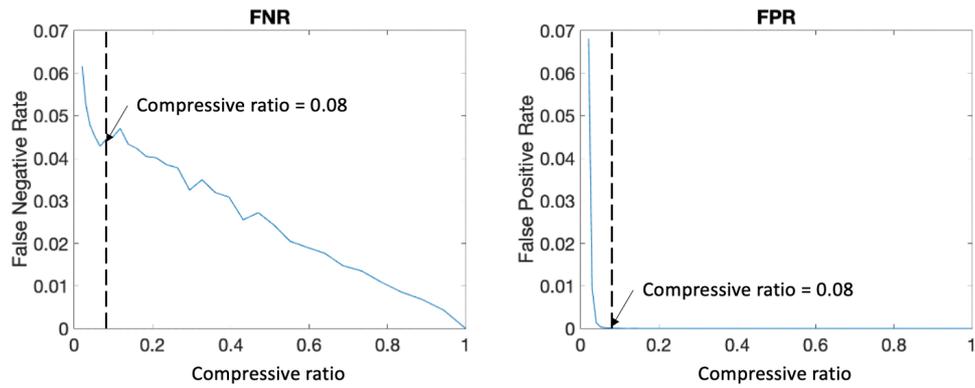

Figure 6. Average FNR and FPR

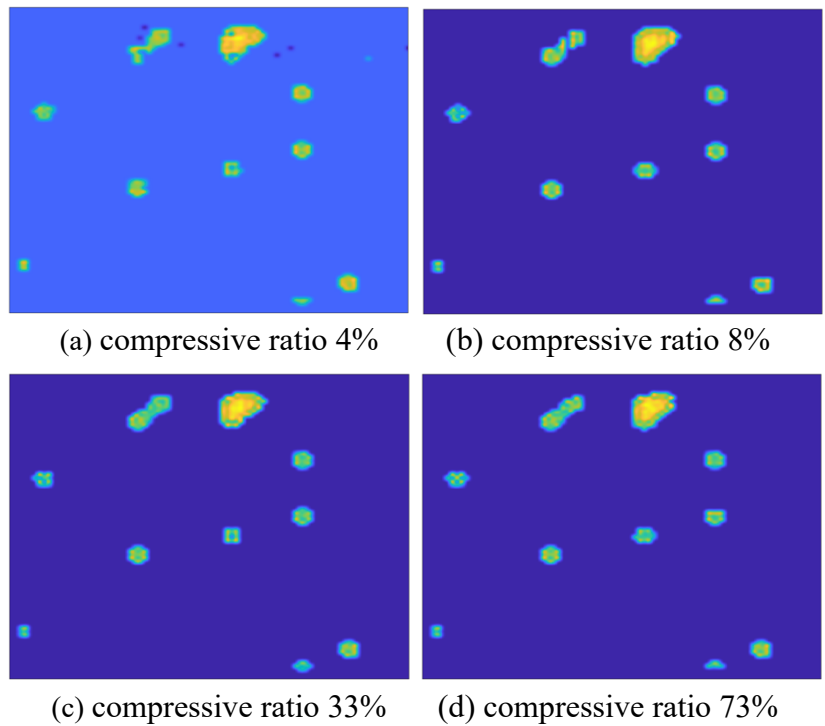

(a) compressive ratio 4%   (b) compressive ratio 8%

(c) compressive ratio 33%   (d) compressive ratio 73%

Figure 7. The recovered sparse signal components

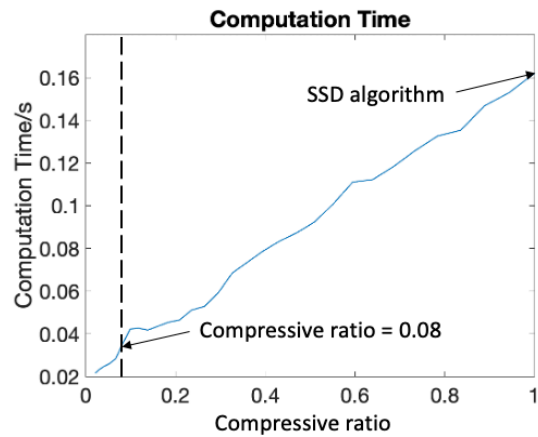

Figure 8. The computation time.

### 3.2.2 Noisy case

The signal-to-noise ratio $\mu/\epsilon \in [4, 40]$ is studied to evaluate the robustness of the algorithm. The 3-D plot of FPR, signal-to-noise ratio, and compressive ratio are shown in Figure 9.

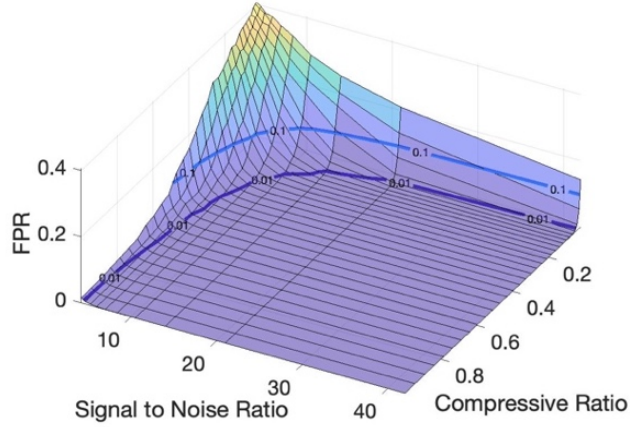

Figure 9. The contour plot of FPR, signal to noise ratio, and compressive ratio.

The purple line indicates the equipotential line of FPR = 0.01. We can see that with the increase of signal-to-noise ratio, we are allowed to use a less compressive ratio to achieve satisfactory decomposition results. The majority of the area lies below the equipotential line of FPR = 0.01, which means the proposed algorithm is robust.

## 4 Case Study

In this section, we use three real cases to demonstrate the effectiveness of the proposed CSSD /KronCSSD framework. For comparison, we also apply the SSD method in each case study. The compressive ratio ($c$) and average computational time ($t$) for a single image are reported in Table 1. A significant transmission bandwidth reduction and computation boost can be observed with negligible performance degradation (Figures 10, 11 and 12), compared to the vanilla SSD algorithm.

Table 1. Comparison between KronCSSD and SSD

|          | Surface defect | | Solar flare | | Indentation | |
|----------|---|---|---|---|---|---|
|          | $c$ | $t$/s | $c$ | $t$/s | $c$ | $t$/s |
| KronCSSD | 54% | 0.073 | 22% | 0.034 | 48% | 0.053 |
| SSD      | 1   | 0.135 | 1   | 0.086 | 1   | 0.094 |

### 4.1 Surface defect detection in steel rolling process

As mentioned in the introduction, vision sensors collect high-resolution images of the product surface with a high data acquisition rate in the rolling processes. This poses a challenge in data storage, transmission, and processing. One sample image of size $128 \times 512$ with typical defects is shown in Figure 10 (a). The black scratches shown in the red block are surface defects. For a detailed description, the readers are encouraged to refer to Yan et al. (2018). The detected anomalies are shown in Figure 10

(b) and (c) by using the proposed KronCSSD algorithm and the SSD algorithm respectively. The data set has 100 images, and more example results can be found in Appendix I.

The KronCSSD method achieves a similar anomaly detection performance with 54% bandwidth and 1.8 times faster. By adopting the KronCSSD method we can (i) achieve a faster anomaly detection and thus reduce the loss through a timely intervention of the manufacturing process; (ii) keep the manufacturing inspection information for a longer time period with the same storage capability which is important for root cause analysis.

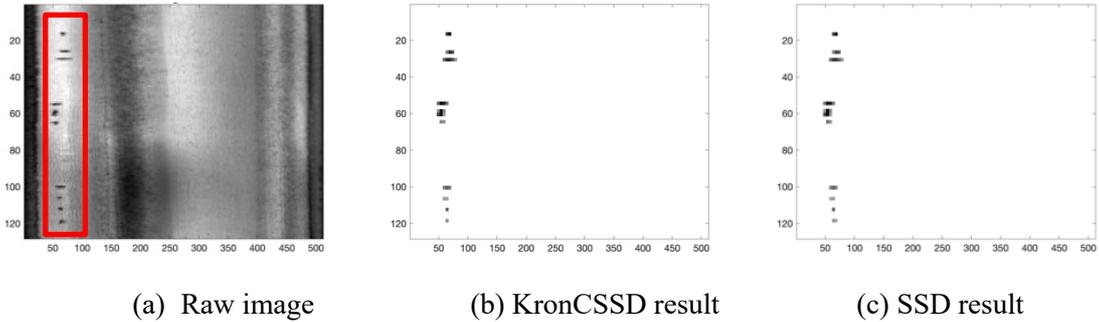

(a) Raw image      (b) KronCSSD result      (c) SSD result

Figure 10. Steel rolling images and detected defects.

## *4.2 Solar flare detection*

Another important application is solar flare detection from satellite images. A solar flare is defined as a sudden, transient, and intense variation in brightness over the Sun's surface. It has a significant influence on radio communication on the earth. Each second, thousands of high-resolution images are captured by a satellite, which poses a big challenge to real-time data transmission and processing (Yan et al., 2018). The data set has 300 images, and more example images can be found in Appendix I.

One sample image of size $232 \times 292$ with a typical solar flare is shown in Figure 11 (a), where the yellowish bright region is the solar flare. The detected anomalies are shown in Figure 11 (b) and (c) by using the proposed KronCSSD and the SSD algorithm respectively. The KronCSSD method achieves a similar solar flare detection performance with the SSD method but with 22% bandwidth and 2.5 times faster. Notice that the decomposed images can also be used for downstream tasks such as control charts and so on, which is beyond the scope of this paper.

By adopting the KronCSSD method we can improve the transmission rate under the same transmission bandwidth and thus achieve almost five times faster solar flare detection, which is of vital importance for protecting radio communications, power grids and navigation systems.

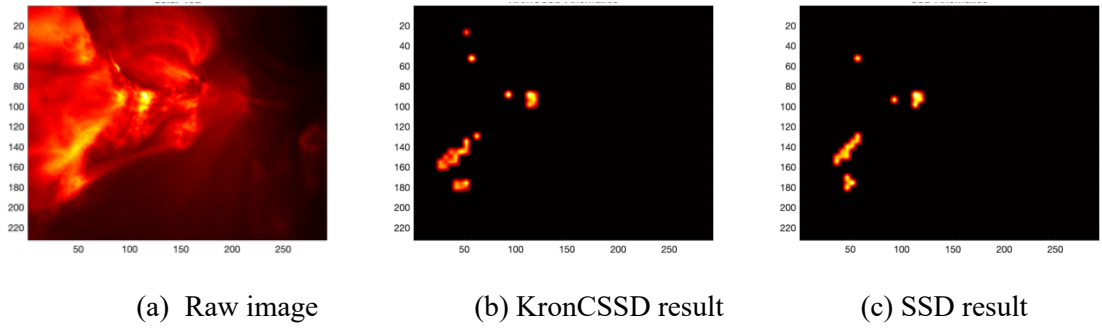

(a) Raw image　　　　(b) KronCSSD result　　　　(c) SSD result

Figure 11. Solar activity images and detected solar flare.

## 4.3  Silicon surface indentation detection

The stress map of size $90 \times 550$ of a silicon surface laminate with surface indentation is shown in Figure 12, where clusters of high-stress areas indicate the surface indentation (Yan et al., 2017). We aim to detect those high-stress areas. The detected anomalies are shown in Figure 12 (b) and (c) respectively. We can see that the KronCSSD method achieves a similar detection performance with the SSD method but with 40% bandwidth is used and 1.8 times faster which significantly reduces the storage and transmission cost and improves the processing speed.

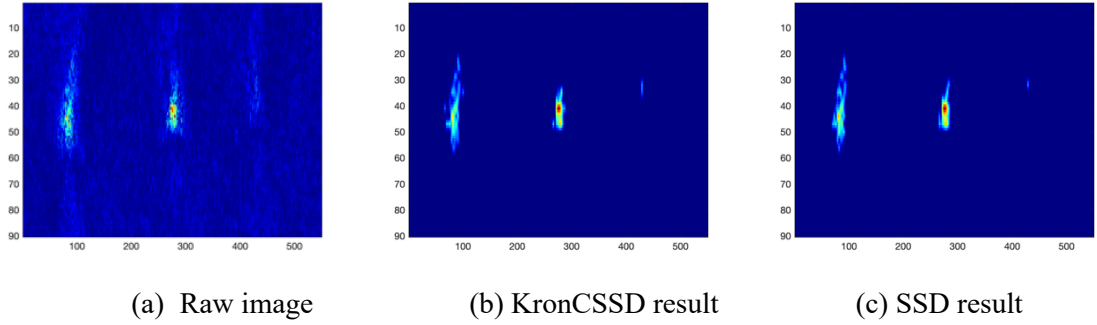

(a) Raw image　　　　(b) KronCSSD result　　　　(c) SSD result

Figure 12. Silicon stress map and detected indentation.

## 5  Conclusion

In this paper, we proposed a CSSD framework for efficient data acquisition, transmission, and processing for sparse anomaly detection in smooth backgrounds. To further enhance its computational efficiency, a KronCSSD framework is proposed for tensor data.

The contributions of this work are in two folds: (i) Theoretically, we showed the feasibility of combining compressive sensing and smooth sparse decomposition. This enables adopting a compressive data acquisition approach. (ii) Practically, the proposed framework is compatible with many existing decomposition-based anomaly detection algorithms, such as SSD, ST-SSD, and so on, which achieves a significant cost reduction in both sensing, storage, and transmission and a boost in speed but with negligible loss in their performance.

In this article, we use a simulation study to demonstrate the effectiveness and robustness of the proposed CSSD/ KronCSSD framework. Three case studies across different applications demonstrate the versatility of the proposed framework. The authors believe that the CSSD/ KronCSSD framework can be applied in a wider range of applications towards more efficient data acquisition, transmission, and processing.

Further studies on the theoretical properties of the proposed KronCSSD can be a future direction.

# 6 Acknowledgment

This work is partially support by NSF Division of Engineering Education and Centers (EEC); Award Number: 2052714.

# 7 Appendices

**Appendix A.  Proof of Theorem 2.1.**

We prove Theorem 2.1 by contradiction: Assume there exists two different decompositions for the same smooth plus sparse signal $y$, i.e., $y = m_1 + a_1 = m_2 + a_2$, where $m_1 \neq m_2$ and $a_1 \neq a_2$. Then,

$$m_1 - m_2 = -a_1 + a_2 \tag{5}$$

Since $m_1 \neq m_2$, we can normalize both side by $\|m_1 - m_2\|_2$ and denote $\widetilde{m} = (m_1 - m_2)/\|m_1 - m_2\|_2$ and $\widetilde{a} = (-a_1 + a_2)/\|m_1 - m_2\|_2$. Notice that $\widetilde{m}$ is in the column space of $B$ which is spanned by columns of the $U$ (recall that $B = U\Sigma V^T$), i.e., $\widetilde{m} = Ux$, where $x$ is the coefficient vector, $x \in \mathbb{R}^r$. Since $\|\widetilde{m}\|_2 = 1$, we have $\|x\|_2 = 1$. We can bound each element in $\widetilde{m}$ as follows:

$$|\widetilde{m}_i| = e_i^T U x \leq \|e_i^T U\|_2 \|x\|_2 \leq \max_{i \in \{1,\dots,r\}} \|U^T e_i\|_2 \leq \sqrt{\frac{\mu(B)r}{n}} < \sqrt{\frac{1}{2ls}}, \quad \forall i \in \{1, \dots, n\}.$$

According to Definition 2.1, $\|a_1\|_0 \leq ls$ and $\|a_2\|_0 \leq ls$. Therefore, $\|\widetilde{a}\|_0 \leq 2ls$. Moreover, according to Eq. (5), we conclude that $\|\widetilde{m}\|_0 \leq 2ls$. Denote the support of $\widetilde{m}$ as $T_{\widetilde{m}}$, we have,

$\|\widetilde{m}\|_2 = \sqrt{\sum_{i \in T_{\widetilde{m}}} |\widetilde{m}_i|^2} < 1$ which is a contradiction. ∎

**Appendix B.  Proof of Theorem 2.2.**

We prove the Theorem 2.2 with contradiction: Assume there exists another vector $y = B\theta + B_a\theta_a \in MS_{r,s,\mu,l}$ such that $Ay = b$ and $y \neq y_0$. Then, $z = y - y_0 = B(\theta - \theta_0) + B_a(\theta_a - \theta_{a0})$ is a nonzero vector. Since $\|\theta_a - \theta_{a0}\|_1 \leq \|\theta_a\|_1 + \|\theta_{a0}\|_1 \leq 2s$, by Definition 2.3, we have that $z \in MS_{r,2s,\mu,l}$. Therefore, $0 = \|Az\|_2 \geq (1 - \delta_{r,2s,\mu,l})\|z\|_2 > 0$, which is a contradiction. Notice that the proof is inspired by the proof of Lemma 3.1 in Candes and Tao (2005). ∎

**Appendix C.  Proof of Theorem 2.3.**

This proof is inspired by Baraniuk et al. (2008) and Tanner and Vary (2020).

We will first derive the RIC for a fixed subspace $MS_{r,T,\mu,l}$ of $MS_{r,s,\mu,l}$ when $\theta_a$ is restricted in a fixed subspace $T$ with the fixed support such that the number of non-zero elements is $s$.

$MS_{r,T,\mu,l} = \{y \in \mathbb{R}^n \mid y = B\theta + B_a\theta_a, B_a \in \mathbb{R}^{n \times q}, l(B_a) = l, \theta_a \in T, B \in \mathbb{R}^{n \times r}, \mu(B) = \mu < n(2rsl)^{-1}, \theta \in \mathbb{R}^r\}$.

Then, we use a covering argument that counts over all possible sparse subspaces **T** with support less than or equal to $s$. Finally, we can derive the RIC for $MS_{r,s,\mu,l}$.

The following lemma describes the RIC for a fixed subspace $MS_{r,T,\mu,l}$ and is proved in Appendix E.

**Lemma 6.1.** *RIC for a fixed subspace $MS_{r,T,\mu,l}$. Let $A \in \mathbb{R}^{p \times n}$ be a matrix from the families described in Definition 3.5. Further, assume $\mu < n(2rsl)^{-1}$ and basis matrix $B_a$ for sparse signal component satisfies the RIP with RIC $\delta_{B_a,s} \in (0,1)$, i.e., $\delta_{B_a,s}$ is the smallest positive constant such that*

$$(1 - \delta_{B_a,s})\|\theta_a\|_2 \le \|B_a\theta_a\|_2 \le (1 + \delta_{B_a,s})\|\theta_a\|_2, \quad \forall\, \theta_a \in \{\theta_a \in \mathbb{R}^q \mid \|\theta_a\|_0 \le s\}.$$

*For a given $\delta \in (0,1)$, there exists a constant $c_0 > 0$ depending only on $\delta$, such that the RIC for $MS_{r,T,\mu,l}$ is upper bounded by $\delta$ with the probability of at least $1 - 2\left(\frac{24}{\delta}\tau_1\right)^r \left(\frac{24}{\delta}\tau_0\right)^s e^{-pc_0(\delta/2)}$, where $\eta = \sqrt{\frac{\mu rsl}{n}}$, $\tau_0 = \frac{1}{\sqrt{(1-\delta_{B_a,s})(1-\eta^2)}}$, $\tau_1 = \|B^\dagger\|_2 \left(1 + \frac{1}{\sqrt{1-\eta^2}}\right)$, and $B^\dagger = (B^T B)^{-1} B^T$.* ■

Notice that for a fixed subspace $MS_{r,T,\mu,l}$, the RIP will fail with probability less than or equal to $2\left(\frac{24}{\delta}\tau_1\right)^r \left(\frac{24}{\delta}\tau_0\right)^s e^{-pc_0(\delta/2)}$. Since there are $\binom{n}{s} \le \left(\frac{en}{s}\right)^s$ such subspaces, the probability to fail for $MS_{r,s,\mu,l}$, which is a combination of those $\binom{n}{s}$ subspaces, will be less than or equal to

$$\binom{n}{s} 2\left(\frac{24}{\delta}\tau_1\right)^r \left(\frac{24}{\delta}\tau_0\right)^s e^{-pc_0\left(\frac{\delta}{2}\right)} \le \exp\left(-c_0\left(\frac{\delta}{2}\right)p + \ln 2 + r\ln\frac{24}{\delta}\tau_1 + s\left(1 + \ln\frac{24}{\delta}\tau_0 + \ln\frac{n}{s}\right)\right).$$

Then, for any give $\delta$, there exist $c_1, c_2 > 0$, such that the probability to fail for $MS_{r,s,\mu,l}$ is less than or equal to $\exp(-c_1 p)$, provided that $p \ge c_2 \left(\ln 2 + r\ln\frac{24}{\delta}\tau_1 + s\left(1 + \ln\frac{24}{\delta}\tau_0 + \ln\frac{n}{s}\right)\right)$, where $c_2 = \left[c_0\left(\frac{\delta}{2}\right) - c_1\right]^{-1}$. This finishes the proof. ■

**Appendix D. Proof of Theorem 2.4.**

The proof of Theorem 2.4 is inspired by Candes et al. (2006) and Tanner and Vary (2020). Assume that in Problem (1), $\epsilon_1$ is properly chosen such that Problem (1) is feasible. In the following discussion, we will use $(\cdot)^*$ to denote the optimal solution of Problem (1) and $(\cdot)_0$ to denote the signal we wish to recover. Let $R = X^* - X_0 = R^m + R^a$ where $R^m = m - m_0 = B\theta^* - B\theta_0$ and $R^a = B_a\theta_a^* - B_a\theta_{a0}$ are the residual of the smooth and sparse signal component, respectively.

Let $h = \theta_a^* - \theta_{a0} = h_{T_0} + h_{T_0^c}$, where $T_0$ is the support of $\theta_{a0}$ and $h_{T_0}$ denotes the projection of $h$ onto $T_0$ such that

$$h_{T_0}(t) = \begin{cases} t, & \text{if } t \in T_0 \\ 0, & \text{otherwise} \end{cases},$$

and $T_0^c$ denotes the complementary set of $T_0$.

Since $\boldsymbol{\theta}_{a0}$ is feasible and $\boldsymbol{\theta}_a^*$ is the optimal solution of Problem (1), we must have $\|\boldsymbol{\theta}_a^*\|_1 \leq \|\boldsymbol{\theta}_{a0}\|_1$, which is equivalent to $\|\boldsymbol{\theta}_{a0} + \boldsymbol{h}_{T_0} + \boldsymbol{h}_{T_0^c}\|_1 \leq \|\boldsymbol{\theta}_{a0}\|_1$. Since $T_0$ and $T_0^c$ are complementary to each other, we have $\|\boldsymbol{\theta}_{a0} + \boldsymbol{h}_{T_0}\|_1 + \|\boldsymbol{h}_{T_0^c}\|_1 \leq \|\boldsymbol{\theta}_{a0}\|_1$. Since $\|\boldsymbol{\theta}_{a0} + \boldsymbol{h}_{T_0}\|_1 \geq \|\boldsymbol{\theta}_{a0}\|_1 - \|\boldsymbol{h}_{T_0}\|_1$, we have $\|\boldsymbol{\theta}_{a0}\|_1 - \|\boldsymbol{h}_{T_0}\|_1 + \|\boldsymbol{h}_{T_0^c}\|_1 \leq \|\boldsymbol{\theta}_{a0}\|_1$. Hence $\|\boldsymbol{h}_{T_0^c}\|_1 \leq \|\boldsymbol{h}_{T_0}\|_1$. Since $\|\boldsymbol{h}_{T_0}\|_1 \leq \sqrt{s}\|\boldsymbol{h}_{T_0}\|_2$, we have

$$\|\boldsymbol{h}_{T_0^c}\|_1 \leq \sqrt{s}\|\boldsymbol{h}_{T_0}\|_2 \tag{6}$$

Similar to Candes et al. (2006), we order the elements of $T_0^c$ in decreasing order of their magnitude and enumerate $T_0^c$ as $v_1, \ldots, v_{n-|T_0|}$. Then, $T_0^c$ is divided into subsets $T_i^c$ of size $M$, where

$$T_i^c = \{v_j : (i-1)M \leq j \leq iM\}.$$

Let $\boldsymbol{h}_{T_i^c}$ be the projection of $\boldsymbol{h}$ onto $T_i^c$, we have

$$\|\boldsymbol{h}_{T_i^c}\|_0 \leq M, \quad \forall i \geq 1$$

$$T_i^c \cap T_j^c = \emptyset, \quad \forall i \neq j$$

$$\|\boldsymbol{h}_{T_{i+1}^c}\|_2 \leq \frac{1}{\sqrt{M}}\|\boldsymbol{h}_{T_i^c}\|_1, \quad \forall i \geq 1 \tag{7}$$

where the last inequality comes from the fact that $T_0^c$ is in decreasing order, such that

$$\left|\boldsymbol{h}_{T_{i+1}^c}\right|_{(v)} \leq \frac{1}{M} \sum_{j \in T_i^c} \left|\boldsymbol{h}_{T_i^c}\right|_{(j)}. \quad \forall v \in T_{i+1}^c$$

Define $\boldsymbol{R}_{T_i^c}^a = \boldsymbol{B}_a \boldsymbol{h}_{T_i^c}$, $\boldsymbol{R}_{T_0}^a = \boldsymbol{B}_a \boldsymbol{h}_{T_0}$ and combine Eq. (6) and Eq. (7), we have

$$\sum_{j \geq 2} \|\boldsymbol{R}_{T_j^c}^a\|_2 \leq \sum_{j \geq 2} \sqrt{1 + \delta_{B_a, M}} \|\boldsymbol{h}_{T_j^c}\|_2 \leq_{(a)} \sum_{j \geq 1} \frac{\sqrt{1 + \delta_{B_a, M}} \|\boldsymbol{h}_{T_j^c}\|_1}{\sqrt{M}} = \frac{\sqrt{1 + \delta_{B_a, M}} \|\boldsymbol{h}_{T_0^c}\|_1}{\sqrt{M}}$$

$$\leq_{(b)} \frac{\sqrt{s}\sqrt{1 + \delta_{B_a, M}}\|\boldsymbol{h}_{T_0}\|_2}{\sqrt{M}} \leq \frac{\sqrt{s}\sqrt{1 + \delta_{B_a, M}}\|\boldsymbol{R}_{T_0}^a\|_2}{\sqrt{M}\sqrt{1 - \delta_{B_a, M}}},$$

where (a) follows Eq.(7), (b) follows Eq. (6) and the RIP property of $\boldsymbol{B}_a$ is used since $\|\boldsymbol{h}_{T_j^c}\|_0 \leq M$.

Denote $\gamma = \sqrt{\frac{1+\delta_{B_a, M+s}}{1-\delta_{B_a, M+s}}}$ (A tighter bound can be achieved by using $\sqrt{\frac{1+\delta_{B_a, M}}{1-\delta_{B_a, M}}}$. However, we adopt $\delta_{B_a, M+s}$ instead of $\delta_{B_a, M}$ for simplicity in the following proof), we have

$$\sum_{j \geq 2} \|\boldsymbol{R}_{T_j^c}^a\|_2 \leq \sqrt{\frac{s}{M}} \gamma \|\boldsymbol{R}_{T_0}^a\|_2 \tag{8}$$

Next, we derive the bound for $\boldsymbol{R}^m$ and $\boldsymbol{R}^a$ respectively.

Bound for $\boldsymbol{R}^m$:

$$\|A\boldsymbol{R}^m\|_2^2 = |\langle A\boldsymbol{R}^m, A(\boldsymbol{R} - \boldsymbol{R}^a)\rangle| \tag{9}$$
$$= |\langle A\boldsymbol{R}^m, A(\boldsymbol{R} - \boldsymbol{R}^a)\rangle|$$

$$= |\langle AR^m, AR \rangle + \langle AR^m, -AR^a \rangle|$$
$$\leq |\langle AR^m, AR \rangle| + |\langle AR^m, -AR^a \rangle|$$
$$= |\langle AR^m, AR \rangle| + \left|\langle AR^m, -A\left(R^a_{T_0} + R^a_{T_1^c} + \sum_{j\geq 2} R^a_{T_j^c}\right)\rangle\right|$$
$$\leq |\langle AR^m, AR \rangle| + \left|\langle AR^m, A\left(R^a_{T_0} + R^a_{T_1^c}\right)\rangle\right| + \sum_{j\geq 2}\left|\langle AR^m, AR^a_{T_j^c}\rangle\right|.$$

In the following discussion, we will bound those terms respectively. According to Cauchy-Schwarz inequality, the first term can be bounded as

$$|\langle AR^m, AR \rangle| \leq \|AR^m\|_2 \|AR\|_2 \leq \sqrt{1+\delta_{r,s,\mu,l}}\epsilon_1 \|R^m\|_2, \tag{10}$$

where the last inequality comes from the RIP property and the first constraint in Problem (1).

The third term can be bounded as follows: denote $z_1 = R^m/\|R^m\|_2$ and $z_2 = R^a_{T_j^c}/\left\|R^a_{T_j^c}\right\|_2$, we have

$$\frac{\left|\langle AR^m, AR^a_{T_j^c}\rangle\right|}{\|R^m\|_2 \left\|R^a_{T_j^c}\right\|_2} = |\langle Az_1, Az_2\rangle| = \frac{1}{4}\left|\|A(z_1+z_2)\|_2^2 - \|A(z_1-z_2)\|_2^2\right|$$

$$\leq_{(a)} \frac{1}{4}\max\begin{Bmatrix}|(1+\delta_{r,M,\mu,l})\|z_1+z_2\|_2^2 - (1-\delta_{r,M,\mu,l})\|z_1-z_2\|_2^2|,\\ |(1+\delta_{r,M,\mu,l})\|z_1-z_2\|_2^2 - (1-\delta_{r,M,\mu,l})\|z_1+z_2\|_2^2|\end{Bmatrix}$$

$$= |\delta_{r,M,\mu,l} + \langle z_1, z_2\rangle| \leq_{(b)} \delta_{r,M,\mu,l} + \frac{\eta_1}{1-\eta_1^2},$$

where $\eta_1 = \sqrt{\frac{\mu rMl}{n}}$; inequality (a) follows the RIP property since $z_1 + z_2 \in MS_{r,M,\mu,l}$ and $z_1 - z_2 \in MS_{r,M,\mu,l}$; inequality (b) follows Eq. (28) in the proof of Lemma 6.2 and $\|z_1\|_2 = \|z_2\|_2 = 1$, $\langle z_1, z_2\rangle \leq \frac{\eta_1}{1-\eta_1^2}\|z_1\|_2\|z_2\|_2 = \frac{\eta_1}{1-\eta_1^2}$, provided that $M \leq 2s$.

Therefore,

$$\left|\langle AR^m, AR^a_{T_j^c}\rangle\right| \leq \left(\delta_{r,M,\mu,l} + \frac{\eta_1}{1-\eta_1^2}\right)\|R^m\|_2 \left\|R^a_{T_j^c}\right\|_2. \tag{11}$$

Similarly, the second term can be bounded as

$$\left|\langle AR^m, A\left(R^a_{T_0} + R^a_{T_1^c}\right)\rangle\right| \leq \left(\delta_{r,M+s,\mu,l} + \frac{\eta_2}{1-\eta_2^2}\right)\|R^m\|_2 \left\|R^a_{T_0} + R^a_{T_1^c}\right\|_2, \tag{12}$$

where $\eta_2 = \sqrt{\frac{\mu r(M+s)l}{n}}$, provided that $M \leq s$.

Plugging Eq. (10), Eq. (11) and Eq. (12) into Eq. (9), we have

$$\|AR^m\|_2^2 \leq \|R^m\|_2 \left(\sqrt{1+\delta_{r,s,\mu,l}}\epsilon_1 + \left(\delta_{r,M+s,\mu,l} + \frac{\eta_2}{1-\eta_2^2}\right)\left\|R^a_{T_0} + R^a_{T_1^c}\right\|_2 + \left(\delta_{r,M,\mu,l} + \frac{\eta_1}{1-\eta_1^2}\right)\left\|R^a_{T_j^c}\right\|_2\right)$$

$$\leq_{(a)} \|R^m\|_2 \left(\sqrt{1+\delta_{r,s,\mu,l}}\epsilon_1 + \left(\delta_{r,M+s,\mu,l} + \frac{\eta_2}{1-\eta_2^2}\right)\left\|R^a_{T_0} + R^a_{T_1^c}\right\|_2 + \left(\delta_{r,M,\mu,l} + \frac{\eta_1}{1-\eta_1^2}\right)\sqrt{\frac{s}{M}}\gamma\|R^a_{T_0}\|_2\right),$$

where inequality (a) follows Eq. (8).

According to the RIP property, we have

$$(1 - \delta_{r,s,\mu,l}) \|R^m\|_2^2$$

$$\leq \|R^m\|_2 \left( \sqrt{1 + \delta_{r,s,\mu,l}} \epsilon_1 + \left( \delta_{r,M+s,\mu,l} + \frac{\eta_2}{1-\eta_2^2} \right) \left\| R_{T_0}^a + R_{T_1^c}^a \right\|_2 + \left( \delta_{r,M,\mu,l} + \frac{\eta_1}{1-\eta_1^2} \right) \sqrt{\frac{s}{M}} \gamma \|R_{T_0}^a\|_2 \right).$$

Consequently, we have

$$\|R^m\|_2 \leq \frac{\sqrt{1 + \delta_{r,s,\mu,l}} \epsilon_1 + \left( \left( \delta_{r,M,\mu,l} + \frac{\eta_1}{1 - \eta_1^2} \right) \sqrt{\frac{s}{M}} \gamma^2 + \left( \delta_{r,M+s,\mu,l} + \frac{\eta_2}{1 - \eta_2^2} \right) \right) \left\| R_{T_0}^a + R_{T_1^c}^a \right\|_2}{(1 - \delta_{r,s,\mu,l})}. \tag{13}$$

where the inequality follows from:

$$\|R_{T_0}^a\|_2 = \|B_a h_{T_0}\|_2 \leq_{(a)} \sqrt{1 + \delta_{B_a,s}} \|h_{T_0}\|_2 \leq_{(b)} \sqrt{1 + \delta_{B_a,s}} \|h_{T_0} + h_{T_1^c}\|_2$$

$$\leq_{(c)} \sqrt{\frac{1 + \delta_{B_a,s}}{1 - \delta_{B_a,M+s}}} \|B_a(h_{T_0} + h_{T_1^c})\|_2 \leq \gamma \left\| R_{T_0}^a + R_{T_1^c}^a \right\|_2, \tag{14}$$

where inequalities (a) and (c) follow the RIP property of $B_a$ and inequality (b) follows that $T_0 \cap T_1^c = \emptyset$.

Bound for $R^a$:

$$\left\| A \left( R_{T_0}^a + R_{T_1^c}^a \right) \right\|_2^2 = \left| \langle A \left( R_{T_0}^a + R_{T_1^c}^a \right), A \left( R_{T_0}^a + R_{T_1^c}^a - R + R \right) \rangle \right|$$

$$= \left| \langle A \left( R_{T_0}^a + R_{T_1^c}^a \right), AR \rangle \right| + \left| \langle A \left( R_{T_0}^a + R_{T_1^c}^a \right), -A \left( R^m + \sum_{j \geq 2} R_{T_j^c}^a \right) \rangle \right| \tag{15}$$

$$\leq \left| \langle A \left( R_{T_0}^a + R_{T_1^c}^a \right), AR \rangle \right| + \left| \langle AR^m, A \left( R_{T_0}^a + R_{T_1^c}^a \right) \rangle \right| + \sum_{j \geq 2} \left| \langle A \left( R_{T_0}^a + R_{T_1^c}^a \right), AR_{T_j^c}^a \rangle \right|$$

In the following discussion, we will bound those terms respectively. According to Cauchy-Schwarz inequality, the first term can be bounded as

$$\left| \langle A \left( R_{T_0}^a + R_{T_1^c}^a \right), AR \rangle \right| \leq \left\| A \left( R_{T_0}^a + R_{T_1^c}^a \right) \right\|_2 \|AR\|_2 \leq \sqrt{1 + \delta_{r,M+s,\mu,l}} \epsilon_1 \left\| R_{T_0}^a + R_{T_1^c}^a \right\|_2, \tag{16}$$

where the last inequality comes from the RIP property and the first constraint in Problem (1).

The third term can be bounded as follows: denote $z_2 = R_{T_j^c}^a / \left\| R_{T_j^c}^a \right\|_2$, $j \geq 2$, and $z_3 = \left( R_{T_0}^a + R_{T_1^c}^a \right) / \left\| R_{T_0}^a + R_{T_1^c}^a \right\|_2$, we have

$$\frac{\left| \langle A \left( R_{T_0}^a + R_{T_1^c}^a \right), AR_{T_j^c}^a \rangle \right|}{\left\| R_{T_0}^a + R_{T_1^c}^a \right\|_2 \left\| R_{T_j^c}^a \right\|_2} = |\langle Az_3, Az_2 \rangle|$$

$$= \frac{1}{4} \left| \|A(z_3 + z_2)\|_2^2 - \|A(z_3 - z_2)\|_2^2 \right|$$

$$\leq_{(a)} \frac{1}{4} \max \left\{ \begin{array}{l} \left| (1 + \delta_{r,2M+s,\mu,l}) \|z_3 + z_2\|_2^2 - (1 - \delta_{r,2M+s,\mu,l}) \|z_3 - z_2\|_2^2 \right|, \\ \left| (1 + \delta_{r,M+2s,\mu,l}) \|z_3 - z_2\|_2^2 - (1 - \delta_{r,2M+s,\mu,l}) \|z_3 + z_2\|_2^2 \right| \end{array} \right\}$$

$$= |\delta_{r,2M+s,\mu,l} + \langle z_3, z_2 \rangle|$$

$$\leq_{(b)} \delta_{r,2M+s,\mu,l},$$

where inequality (a) follows the RIP property since $z_3 + z_2 \in MS_{r,2M+s,\mu,l}$ and $z_3 - z_2 \in MS_{r,2M+s,\mu,l}$; inequality (b) comes from that $\mathbf{T}_i^c \cap \mathbf{T}_j^c = \emptyset, \ \forall i \neq j$ and $\mathbf{T}_0 \cap \mathbf{T}_i^c = \emptyset, \forall i$.

Therefore,

$$\left|\langle A R^m, A R_{\mathbf{T}_j^c}^a \rangle\right| \leq \delta_{r,2M+s,\mu,l} \left\| R_{\mathbf{T}_0}^a + R_{\mathbf{T}_1^c}^a \right\|_2 \left\| R_{\mathbf{T}_j^c}^a \right\|_2. \tag{17}$$

Plugging Eq. (12), Eq. (16) and Eq. (17) into Eq. (15), we have

$$\left\| A\left(R_{\mathbf{T}_0}^a + R_{\mathbf{T}_1^c}^a\right) \right\|_2^2$$

$$\leq \left\| R_{\mathbf{T}_0}^a + R_{\mathbf{T}_1^c}^a \right\|_2 \left( \sqrt{1 + \delta_{r,M+s,\mu,l}} \epsilon_1 + \left(\delta_{r,M+s,\mu,l} + \frac{\eta_2}{1-\eta_2^2}\right) \|R^m\|_2 + \delta_{r,2M+s,\mu,l} \sum_{j \geq 2} \left\| R_{\mathbf{T}_j^c}^a \right\|_2 \right)$$

$$\leq_{(a)} \left\| R_{\mathbf{T}_0}^a + R_{\mathbf{T}_1^c}^a \right\|_2 \left( \sqrt{1 + \delta_{r,M+s,\mu,l}} \epsilon_1 + \left(\delta_{r,M+s,\mu,l} + \frac{\eta_2}{1-\eta_2^2}\right) \|R^m\|_2 + \delta_{r,2M+s,\mu,l} \sqrt{\frac{s}{M}} \gamma \left\| R_{\mathbf{T}_0}^a \right\|_2 \right)$$

$$\leq_{(b)} \left\| R_{\mathbf{T}_0}^a + R_{\mathbf{T}_1^c}^a \right\|_2 \left( \sqrt{1 + \delta_{r,M+s,\mu,l}} \epsilon_1 + \left(\delta_{r,M+s,\mu,l} + \frac{\eta_2}{1-\eta_2^2}\right) \|R^m\|_2 + \delta_{r,2M+s,\mu,l} \sqrt{\frac{s}{M}} \gamma^2 \left\| R_{\mathbf{T}_0}^a + R_{\mathbf{T}_1^c}^a \right\| \right),$$

where inequalities (a) and (b) follows the same argument as deriving Eq. (13).

According to the RIP property, we have

$$\left(1 - \delta_{r,M+s,\mu,l}\right) \left\| R_{\mathbf{T}_0}^a + R_{\mathbf{T}_1^c}^a \right\|_2^2$$

$$\leq \left\| R_{\mathbf{T}_0}^a + R_{\mathbf{T}_1^c}^a \right\|_2 \left( \sqrt{1 + \delta_{r,M+s,\mu,l}} \epsilon_1 + \left(\delta_{r,M+s,\mu,l} + \frac{\eta_2}{1-\eta_2^2}\right) \|R^m\|_2 + \delta_{r,2M+s,\mu,l} \sqrt{\frac{s}{M}} \gamma^2 \left\| R_{\mathbf{T}_0}^a \right\|_2 \right).$$

Consequently, we have

$$\left\| R_{\mathbf{T}_0}^a + R_{\mathbf{T}_1^c}^a \right\|_2$$

$$\leq \frac{\left( \sqrt{1 + \delta_{r,M+s,\mu,l}} \epsilon_1 + \left(\delta_{r,M+s,\mu,l} + \frac{\eta_2}{1-\eta_2^2}\right) \|R^m\|_2 + \delta_{r,2M+s,\mu,l} \sqrt{\frac{s}{M}} \gamma^2 \left\| R_{\mathbf{T}_0}^a + R_{\mathbf{T}_1^c}^a \right\| \right)}{\left(1 - \delta_{r,M+s,\mu,l}\right)} \tag{18}$$

Notice that Eq. (13) and Eq. (18) still hold if we relax $\delta_{r,M+s,\mu,l}, \delta_{r,s,\mu,l}$ to $\delta_{r,2M+s,\mu,l}$. For simplicity, here we replace $\delta_{r,M+s,\mu,l}, \delta_{r,s,\mu,l}$ with $\delta_{r,2M+s,\mu,l}$ in the following derivation.

Plugging Eq. (13) into Eq. (18), and let $x \equiv \left\| R_{\mathbf{T}_0}^a + R_{\mathbf{T}_1^c}^a \right\|_2, y \equiv \|R^m\|_2$, we have

$$\left(D_1 - \frac{B_1 B_2}{D_2} - C_1\right) x \leq A_1 \epsilon_1 + \frac{B_1}{D_2} A_2 \epsilon_1 \tag{19}$$

where

$$A_1 = \sqrt{1 + \delta_{r,2M+s,\mu,l}}, B_1 = \left(\delta_{r,2M+s,\mu,l} + \frac{\eta_2}{1-\eta_2^2}\right), C_1 = \delta_{r,2M+s,\mu,l} \sqrt{\frac{s}{M}} \gamma^2, D_1 = \left(1 - \delta_{r,2M+s,\mu,l}\right),$$

and

$$A_2 = \sqrt{1 + \delta_{r,2M+s,\mu,l}}, B_2 = \left(\delta_{r,2M+s,\mu,l} + \frac{\eta_1}{1-\eta_1^2}\right) \sqrt{\frac{s}{M}} \gamma^2 + \left(\delta_{r,2M+s,\mu,l} + \frac{\eta_2}{1-\eta_2^2}\right), D_2$$

$$= \left(1 - \delta_{r,2M+s,\mu,l}\right).$$

Here we require $D_1 - \frac{B_1 B_2}{D_2} - C_1 > 0$ in Eq. (19) and let $M = s$, which is

$$1 - \gamma^2 \alpha_1 \alpha_2 - \alpha_2^2 - \big((1 + \alpha_1 + \alpha_2)\gamma^2 + 2\alpha_2 + 2\big)\delta_{r,3s,\mu,l} > 0. \tag{20}$$

Let $a = (1 + \alpha_1 + \alpha_2)\gamma^2 + 2\alpha_2 + 2$ and $c = 1 - \gamma^2 \alpha_1 \alpha_2 - \alpha_2^2$, where $\alpha_1 = \frac{\eta}{1-\eta^2}$, $\alpha_2 = \frac{\sqrt{2}\eta}{1-2\eta^2}$, $\gamma = \sqrt{\frac{1+\delta_{B_a,2s}}{1-\delta_{B_a,2s}}}$, $\eta = \sqrt{\frac{\mu r s l}{n}}$. If $c > 0$, then, there exist a $\delta_{r,3s,\mu,l} > 0$ such that Eq. (20) is valid. Then the denominator $-a\delta_{r,3s,\mu,l} + c > 0 \ \forall \ \delta_{r,3s,\mu,l} \in (0, c/a)$. Consequently,

$$\left\| R^a_{T_0} + R^a_{T_1^c} \right\|_2 \leq \frac{(1 + \alpha_2)\sqrt{1 + \delta_{r,3s,\mu,l}}}{c - a\delta_{r,3s,\mu,l}} \epsilon_1.$$

Notice that from Eq. (6) and Eq. (14), we have

$$\sum_{j \geq 2} \left\| R^a_{T_j^c} \right\|_2 \leq \gamma \left\| R^a_{T_0} \right\|_2 \leq \gamma^2 \left\| R^a_{T_0} + R^a_{T_1^c} \right\|_2$$

Therefore, $\|R^a\|_2 \leq \left\| R^a_{T_0} + R^a_{T_1^c} \right\|_2 + \sum_{j \geq 2} \left\| R^a_{T_j^c} \right\|_2 \leq C_a \epsilon_1$, where

$$C_a = \frac{(1 + \gamma^2)(1 + \alpha_2)\sqrt{1 + \delta_{r,3s,\mu,l}}}{c - a\delta_{r,3s,\mu,l}}.$$

Similarly, we can bound $\|R^m\|_2$ as $\|R^m\|_2 \leq C_m \epsilon_1$, where

$$C_m = \frac{\sqrt{1 + \delta_{r,3s,\mu,l}} + \left(\delta_{r,3s,\mu,l} + \frac{\gamma^2}{1+\gamma^2}\alpha_1 + \frac{1}{1+\gamma^2}\alpha_2\right) C_a}{(1 - \delta_{r,3s,\mu,l})}.$$

**Appendix E. Proof of Lemma 6.1.**

In this section, we will provide the proof for Lemma 6.1. By linearity of the measurement matrix $A$, without loss of generality, it is enough to prove this lemma when $\|y\|_2 = 1$. The proof mainly has two steps. First, the bounds for $\theta_a$ and $\theta$ are derived and a finite set of points to approximate the set $MS_{r,T,\mu,l}$ to any accuracy in norm 2 sense can be found. Then, the concentration inequality can be applied through a union bound. This is a common approach in compressive sensing literature (Baraniuk et al., 2008; Tanner & Vary, 2020).

To derive the upper bounds for $\theta_a$ and $\theta$, we first derive the upper bounds for the signal $m = B\theta$ and $a = B_a \theta_a$, which are given in the following lemma.

**Lemma 6.2.** *The smooth signal component $m$ and sparse signal component $a$ of the signal $y$ in $MS_{r,T,\mu,l}$ with $\mu < \frac{n}{2rsl}$ can be bounded as follows*

$$\|m\|_2 = \|B\theta\|_2 \leq \frac{\|y\|_2}{\sqrt{1-\eta^2}}, \tag{21}$$

$$\|a\|_2 = \|B_a \theta_a\|_2 \leq \frac{\|y\|_2}{\sqrt{1-\eta^2}}, \tag{22}$$

where $\eta = \sqrt{\frac{\mu r s l}{n}}$. ∎

The proof is presented in Appendix F.

According to the RIP for $B_a$, we have

$$\sqrt{(1-\delta_{B_a,s})}\|\boldsymbol{\theta}_a\|_2 \leq \|B_a\boldsymbol{\theta}_a\|_2 \leq \sqrt{(1+\delta_{B_a,s})}\|\boldsymbol{\theta}_a\|_2. \qquad (23)$$

Combine Eq. (22) and Eq. (23), we have

$$\|\boldsymbol{\theta}_a\|_2 \leq \frac{1}{\sqrt{(1-\delta_{B_a,s})}}\|B_a\boldsymbol{\theta}_a\|_2 \leq \frac{\|y\|_2}{\sqrt{(1-\delta_{B_a,s})(1-\eta^2)}} = \frac{1}{\sqrt{(1-\delta_{B_a,s})(1-\eta^2)}}.$$

Denote $\tau_0 = \frac{1}{\sqrt{(1-\delta_{B_a,s})(1-\eta^2)}}$, we have $\|\boldsymbol{\theta}_a\|_2 \leq \tau_0$. Recall that $y = B\boldsymbol{\theta} + B_a\boldsymbol{\theta}_a$, we have $\boldsymbol{\theta} = B^\dagger(y - B_a\boldsymbol{\theta}_a)$, where $B^\dagger = (B^\mathrm{T}B)^{-1}B^\mathrm{T}$. Therefore, according to triangle inequality and Cauchy–Schwarz inequality, we have

$$\|\boldsymbol{\theta}\|_2 \leq \|B^\dagger\|_2(\|y\|_2 + \|B_a\boldsymbol{\theta}_a\|_2) \leq \|B^\dagger\|_2\left(1+\frac{1}{\sqrt{1-\eta^2}}\right).$$

Denote $\tau_1 = \|B^\dagger\|_2\left(1+\frac{1}{\sqrt{1-\eta^2}}\right)$, we have $\|\boldsymbol{\theta}\|_2 \leq \tau_1$.

Since we have derived the bounds for $\boldsymbol{\theta}_a$ and $\boldsymbol{\theta}$, the covering number of $MS_{r,\mathrm{T},\mu,l}$ is given by the following lemma whose proof is in Appendix G.

**Lemma 6.3.** *There exists a set $Q \in MS_{r,\mathrm{T},\mu,l}$, such that for all $y \in MS_{r,\mathrm{T},\mu,l}$, with $\|y\|_2 = 1$ we have $\min_{q \in Q}\|q - y\|_2 \leq \frac{\delta}{4}$, and $|Q| \leq \left(\frac{24}{\delta}\tau_1\right)^r\left(\frac{24}{\delta}\tau_0\right)^s$, where $|Q|$ is its cardinality.* ∎

Next, we will prove the main result by applying the concentration inequality (Definition 3.5. (ii)) with union bound. Let $\epsilon = \delta/2$,

$$\left(1-\frac{\delta}{2}\right)\|q\|_2^2 \leq \|Aq\|_2^2 \leq \left(1+\frac{\delta}{2}\right)\|q\|_2^2 \quad \forall q \in Q, \qquad (24)$$

with probability greater than $1 - 2|Q|e^{-pc_0(\delta/2)}$.

Since $\delta \in (0,1)$, we have $1-\frac{\delta}{2} \leq \sqrt{1-\frac{\delta}{2}}$ and $\sqrt{1+\frac{\delta}{2}} \leq 1+\frac{\delta}{2}$. Then Eq. (24) can be written as

$$\left(1-\frac{\delta}{2}\right)\|q\|_2 \leq \|Aq\|_2 \leq \left(1+\frac{\delta}{2}\right)\|q\|_2 \quad \forall q \in Q, \qquad (25)$$

with probability greater than $1 - 2|Q|e^{-pc_0(\delta/2)}$.

By the triangle inequality, we have

$$\|Ay\|_2 \leq \|A(y-q)\|_2 + \|Aq\|_2. \qquad (26)$$

Define

$$U = \max_{y \in MS_{r,\mathrm{T},\mu,l}, \|y\|_2=1}\|Ay\|_2, \qquad (27)$$

which is attainable since $MS_{r,\mathrm{T},\mu,l}$ is closed.

Combine Eq. (25) and Eq. (26), we have $\forall y \in MS_{r,\mathrm{T},\mu,l}$, with $\|y\|_2 = 1$, there exist a $q \in MS_{r,\mathrm{T},\mu,l}$, such that $\|Ay\|_2 \leq \|A(q-y)\|_2 + \left(1+\frac{\delta}{2}\right)\|q\|_2$, with probability greater than $1 - 2|Q|e^{-pc_0(\delta/2)}$.

Since $q - y \in MS_{r,T,\mu,l}$, if $q - y = 0$, we have $\|Ay\|_2 \leq \left(1 + \frac{\delta}{2}\right)\|q\|_2 = 1 + \frac{\delta}{2}$.

If $q - y \neq 0$, we have $\|Ay\|_2 \leq \left\|A\frac{(q-y)}{\|q-y\|_2}\right\|_2 \|q - y\|_2 + \left(1 + \frac{\delta}{2}\right)\|q\|_2 \leq \frac{\delta}{4}U + 1 + \frac{\delta}{2}$.

Notice that the second inequality comes from Eq.(27) combined with $Q$ being a $\frac{\delta}{4}$ covering of $MS_{r,T,\mu,l}$. In summary, we have $\|Ay\|_2 \leq \frac{\delta}{4}U + 1 + \frac{\delta}{2}$.

Since $U$ is attainable, according to Eq. (27), we have $U \leq \frac{\delta}{4}U + 1 + \frac{\delta}{2}$. Consequently, we have $U \leq 1 + \frac{3}{4-\delta}\delta \leq 1 + \delta$, since $\delta < 1$. Therefore, $\|Ay\|_2 \leq 1 + \delta$, with probability greater than $1 - 2|Q|e^{-pc_0(\delta/2)}$.

Similarly, we can prove that $\|Ay\|_2 \geq 1 - \delta$ with probability greater than $1 - 2|Q|e^{-pc_0(\delta/2)}$.

Finally, according to Lemma 6.3, we have that

$$1 - 2|Q|e^{-pc_0\left(\frac{\delta}{2}\right)} \geq 1 - 2\left(\frac{24}{\delta}\tau_1\right)^r \left(\frac{24}{\delta}\tau_0\right)^s e^{-pc_0\left(\frac{\delta}{2}\right)}.$$

This finishes the proof. ∎

**Appendix F.    Proof of Lemma 6.2.**

To prove the result, we first derive a nontrivial upper bound for the inner produce between $m$ and $a$. Let $B = U\Sigma V^T$ be the reduced SVD of $B$, then

$$|m^T a| = |\theta^T B^T a| = |\theta^T V\Sigma U^T a| = \left|\theta^T V\Sigma U^T \sum_i^n a_i e_i\right| = \left|\theta^T V\Sigma \sum_i^n a_i U^T e_i\right|$$

$$\leq \|\theta^T V\Sigma\|_2 \left\|\sum_i^n a_i U^T e_i\right\|_2$$

$$\leq \|\theta^T V\Sigma\|_2 \sum_i^n |a_i| \|U^T e_i\|_2 \qquad (28)$$

$$\leq \|\theta^T V\Sigma\|_2 \sum_i^n |a_i| \max_{j \in \{1,\ldots r\}} \|U^T e_j\|_2$$

$$\leq_{(a)} \|\theta^T V\Sigma U^T\|_2 \|a\|_1 \sqrt{\frac{\mu r}{n}}$$

$$\leq_{(b)} \sqrt{\frac{\mu rsl}{n}} \|m\|_2 \|a\|_2.$$

where inequality (a) follows $\|\theta^T V\Sigma\|_2 = \|\theta^T V\Sigma U^T\|_2$ since $U^T U = I$ and Definition 2.2. Inequality (b) follows from $\|a\|_1 \leq \sqrt{ls}\|a\|_2$.

Let $\eta = \sqrt{\frac{\mu rsl}{n}}$, since $\mu < \frac{n}{rsl}$, we have $\eta < 1$ and

$$|m^T a| = \frac{|\|y\|_2^2 - \|m\|_2^2 - \|a\|_2^2|}{2} \leq \eta \|m\|_2 \|a\|_2.$$

Therefore, we have
$$\|m\|_2^2 + \|a\|_2^2 - \|y\|_2^2 \leq 2\eta\|m\|_2\|a\|_2.$$

By completing the square, we have
$$(\|m\|_2 + \eta\|a\|_2)^2 + (1-\eta^2)\|a\|_2^2 - \|y\|_2^2 \leq 0.$$

Since $(\|m\|_2 + \eta\|a\|_2)^2 \geq 0$, we have
$$\|a\|_2 \leq \frac{1}{\sqrt{(1-\eta^2)}}\|y\|_2.$$

Similarly, we can derive that
$$\|m\|_2 \leq \frac{1}{\sqrt{(1-\eta^2)}}\|y\|_2.$$

This finishes the proof. ∎

**Appendix G. Proof of Lemma 6.3.**

We first state results for the covering number of a set (Vershynin, 2018): The covering number of a smallest $\epsilon$-net for a unit $l_2$ norm ball in $d$ dimensional space is $(3/\epsilon)^d$.

Let $\mathbf{M} = \{m \in \mathbb{R}^n | m = B\theta, \theta \in \mathbb{R}^r, \|\theta\|_2 \leq \tau_1, \mu(B) = \mu\}$ and $\mathbf{S} = \{a \in \mathbb{R}^n | a = B_a\theta_a, \theta_a \in \mathbf{T}, \|\theta_a\|_2 \leq \tau_0, l(B_a) = l\}$. There exists two finite $\frac{\delta}{8}$ covering sets of $\mathbf{M}$ and $\mathbf{S}$, which are $\mathbf{Q_M} \subseteq \mathbf{M}$ and $\mathbf{Q_S} \subseteq \mathbf{S}$.

For all $q_\mathbf{M} \in \mathbf{Q_M}$, and for all $m \in \mathbf{M}$, we have,
$$\min_{q_\mathbf{M} \in \mathbf{Q_M}} \|m - q_\mathbf{M}\|_2 \leq \frac{\delta}{8};$$

For all $q_\mathbf{S} \in \mathbf{Q_S}$, and for all $a \in \mathbf{S}$, we have,
$$\min_{q_\mathbf{S} \in \mathbf{Q_S}} \|a - q_\mathbf{S}\|_2 \leq \frac{\delta}{8}.$$

According to (i) and (ii), we have $|\mathbf{Q_M}| \leq \left(\frac{24}{\delta}\tau_1\right)^r$ and $|\mathbf{Q_S}| \leq \left(\frac{24}{\delta}\tau_0\right)^s$.

Define $\mathbf{Q_{MS}} = \{q_\mathbf{M} + q_\mathbf{S} | q_\mathbf{M} \in \mathbf{Q_M}, q_\mathbf{S} \in \mathbf{Q_S}\} \subseteq MS_{r,\mathbf{T},\mu,l}$. Then $\forall y \in MS_{r,\mathbf{T},\mu,l}$, there exists a pair $q_\mathbf{MS} = q_\mathbf{M} + q_\mathbf{S} \in MS_{r,\mathbf{T},\mu,l}$, such that
$$\|q_\mathbf{MS} - y\|_2 = \|q_\mathbf{M} - m + q_\mathbf{S} - a\|_2 \leq \|q_\mathbf{M} - m\|_2 + \|q_\mathbf{S} - a\|_2 \leq \frac{\delta}{4}.$$

Therefore, $\mathbf{Q_{MS}}$ is a $\delta/4$ covering of $MS_{r,\mathbf{T},\mu,l}$ and $|\mathbf{Q_{MS}}| \leq \left(\frac{24}{\delta}\tau_1\right)^r \left(\frac{24}{\delta}\tau_0\right)^s$. This finishes the proof. ∎

**Appendix H. Simulation study by varying the magnitude of sparse signals**

We adopt the same simulation data generation procedure as in Section 3.1 while changing the distribution of elements in $\theta_a \in \mathbb{R}^q$ such that the non-zero elements follow i.i.d. normal distribution with mean zero and standard deviation $\sigma_s$ in the range of $\{0.065, 0.125, 0.25, 0.5\}$. The example signals and reconstruction performance are shown in Figure 13.

There are several observations:

1. The log relative error of the smooth component does not change with different magnitudes of $\sigma_s$. This agrees with the theoretical result in Theorem 2.4 where the reconstruction error is bounded by a constant times the noise bound since the noise level is kept the same in all simulations.
2. The log relative error of the sparse component decreases as $\sigma_s$ increases. This also agrees with the theoretical result in Theorem 2.4. Since the log relative error of the sparse component is defined as $\log\|a - \hat{a}\|_2/\|a\|_2$, according to Theorem 2.4, the reconstruction error term can be approximately approximated by $C_a\epsilon_1$, i.e., $\|a - \hat{a}\|_2 \sim C_a\epsilon_1$ where $C_a$ is independent of $\boldsymbol{\theta}_a$. $\|a\|_2 = \|B_a\boldsymbol{\theta}_a\|_2$ increases as the magnitude of elements in $\boldsymbol{\theta}_a$ increases. Therefore, $\log\|a - \hat{a}\|_2/\|a\|_2$ will decrease as $\sigma_s$ increases.
3. The 0.1 threshold (approximately) of the compressive ratio still holds with different magnitudes of sparse signal component.

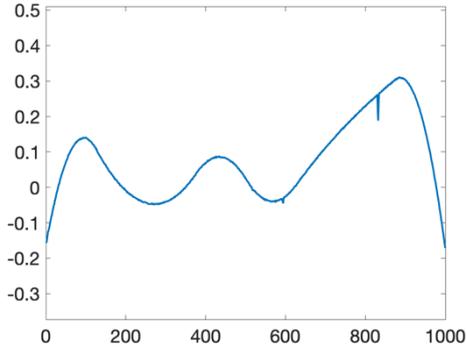

(a) $\sigma_s = 0.065$

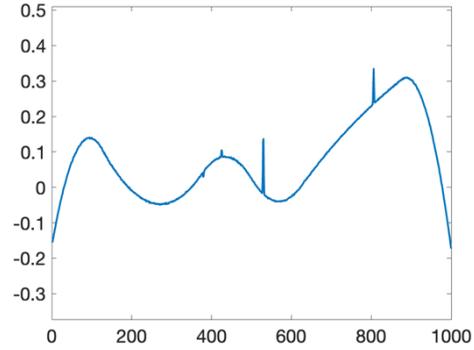

(b) $\sigma_s = 0.125$

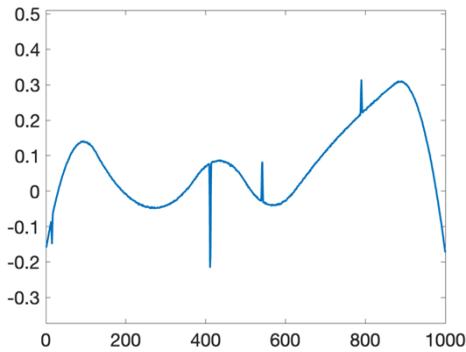

(c) $\sigma_s = 0.25$

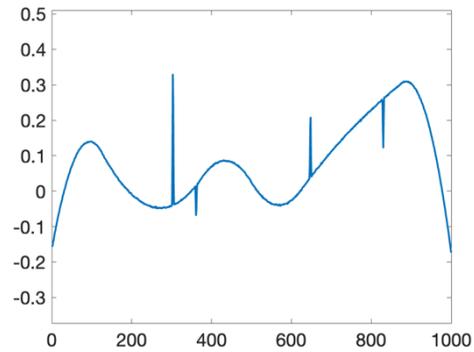

(d) $\sigma_s = 0.5$

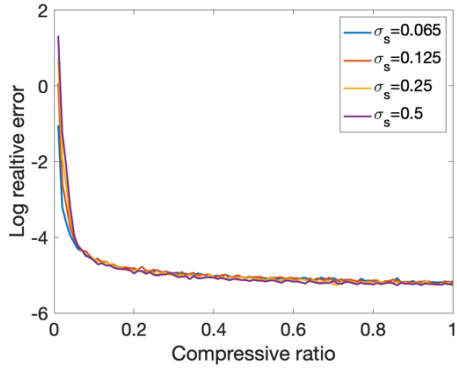
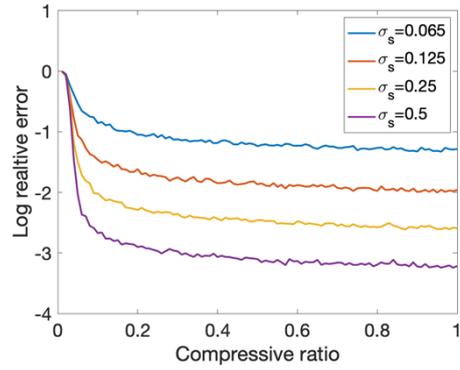

(e) Log relative error for smooth component   (f) Log relative error for sparse component

Figure 13. Simulation study by varying $\sigma_s$. (a)-(d): example signals corresponding to different $\sigma_s$; (e) Log relative error for smooth component; and (f) Log relative error for sparse component.

**Appendix I.    Sample case study images**

I.1. Sample images of case study 4.1

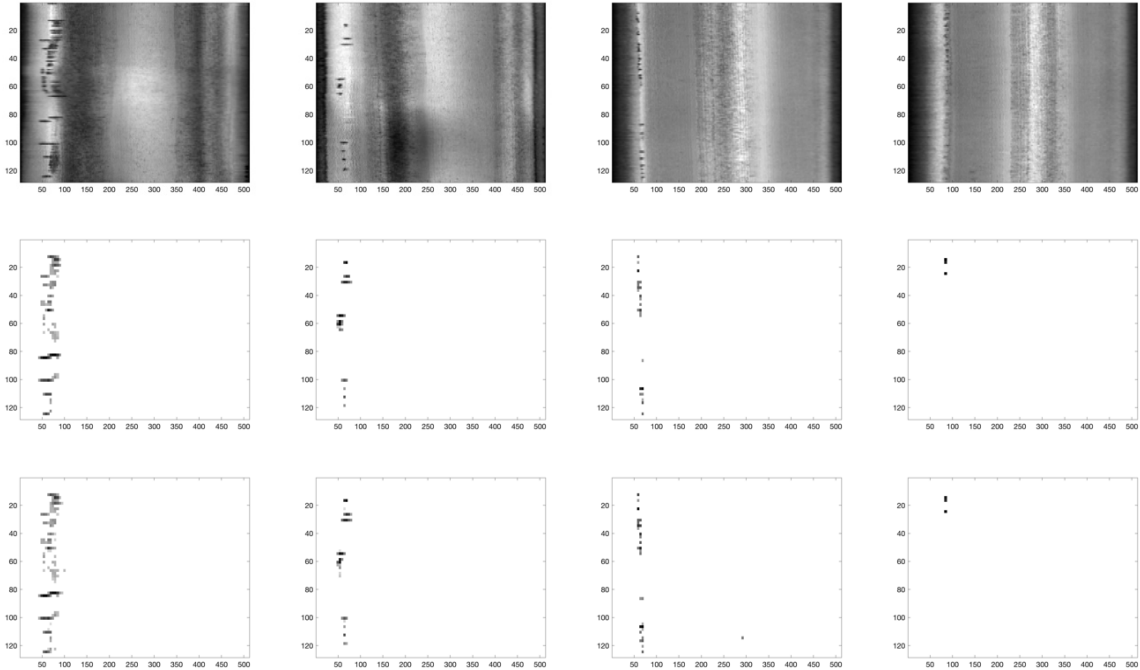

Figure 14. Sample images of surface defect detection in steel rolling process: the first row shows raw images; the second row shows corresponding SSD results; the third row shows KronCSSD results

I.2. Sample images of case study 4.2

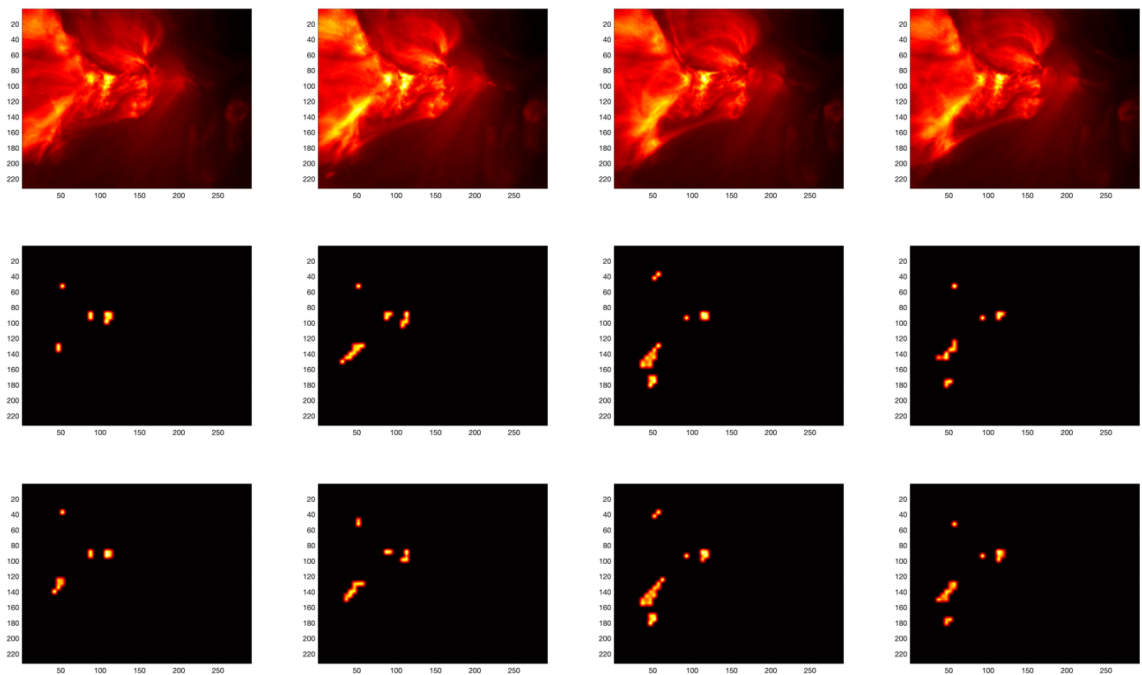

Figure 15. Sample images of solar flare detection: the first row shows raw images; the second row shows corresponding SSD results; the third row shows KronCSSD results